

\input harvmac

\input epsf

\input amssym.def
\input amssym
\baselineskip 14pt
\magnification\magstep1
\parskip 6pt

\font \bigbf=cmbx10 scaled \magstep1

\newdimen\itemindent \itemindent=32pt
\def\textindent#1{\parindent=\itemindent\let\par=\resetpar%
\indent\llap{#1\enspace}\ignorespaces}

\let\oldpar=\par
\def\resetpar{\oldpar\parindent=20pt\let\par=\oldpar}

\font\ninerm=cmr9 \font\ninesy=cmsy9
\font\eightrm=cmr8 \font\sixrm=cmr6
\font\eighti=cmmi8 \font\sixi=cmmi6
\font\eightsy=cmsy8 \font\sixsy=cmsy6
\font\eightbf=cmbx8 \font\sixbf=cmbx6
\font\eightit=cmti8
\def\eightpoint{\def\rm{\fam0\eightrm}
  \textfont0=\eightrm \scriptfont0=\sixrm \scriptscriptfont0=\fiverm
  \textfont1=\eighti  \scriptfont1=\sixi  \scriptscriptfont1=\fivei
  \textfont2=\eightsy \scriptfont2=\sixsy \scriptscriptfont2=\fivesy
  \textfont3=\tenex   \scriptfont3=\tenex \scriptscriptfont3=\tenex
  \textfont\itfam=\eightit  \def\it{\fam\itfam\eightit}%
  \textfont\bffam=\eightbf  \scriptfont\bffam=\sixbf
  \scriptscriptfont\bffam=\fivebf  \def\bf{\fam\bffam\eightbf}%
  \normalbaselineskip=9pt
  \setbox\strutbox=\hbox{\vrule height7pt depth2pt width0pt}%
  \let\big=\eightbig  \normalbaselines\rm}
\catcode`@=11 %
\def\eightbig#1{{\hbox{$\textfont0=\ninerm\textfont2=\ninesy
  \left#1\vbox to6.5pt{}\right.\n@@space$}}}
\def\vfootnote#1{\insert\footins\bgroup\eightpoint
  \interlinepenalty=\interfootnotelinepenalty
  \splittopskip=\ht\strutbox %
  \splitmaxdepth=\dp\strutbox %
  \leftskip=0pt \rightskip=0pt \spaceskip=0pt \xspaceskip=0pt
  \textindent{#1}\footstrut\futurelet\next\fo@t}
\catcode`@=12 %

\def\a{\alpha}
\def\b{\beta}
\def\c{\gamma}
\def\d{\delta}

\def\l{\lambda}
\def\m{\mu}
\def\n{\nu}

\def\r{\rho}

\def\C{\Gamma}
\def\D{\Delta}

\def\S{\Sigma}

\def\pl{\partial}
\def\rta{\rightarrow}

\lref\DH{I.T. Drummond and S. Hathrell, Phys. Rev. D22 (1980) 343.}
\lref\Sone{R.D. Daniels and G.M. Shore, Nucl. Phys. B425 (1994) 634.}
\lref\Stwo{R.D. Daniels and G.M. Shore, Phys. Lett. B367 (1996) 75.}
\lref\Sthree{G.M. Shore, Nucl. Phys. B460 (1996) 379.}
\lref\Sfour{G.M. Shore, Nucl. Phys. B605 (2001) 455.}
\lref\Sfive{G.M. Shore, Nucl. Phys. B633 (2002) 271.}
\lref\Ssix{G.M. Shore, gr-qc/0205042.}
\lref\DN{A.D. Dolgov and I.D. Novikov, Phys. Lett. B442 (1998) 82.}
\lref\Khrip{I.B. Khriplovich, Phys. Lett. B346 (1995) 251.}         
\lref\Konst{M.Yu. Konstantinov, gr-qc/9810019.}
\lref\Ch{S. Chandresekhar, {\it The Mathematical Theory of Black Holes},
Clarendon, Oxford (1985) }
\lref\Bondi{H. Bondi, M.G.J. van der Burg and A.W.K. Metzner, Proc. Roy. Soc.
A269 (1962) 21.}
\lref\Sachs{R.K. Sachs, Proc. Roy. Soc. A270 (1962) 103.}
\lref\LSV{S. Liberati, S. Sonego and M. Visser, Phys. Rev. D63 (201) 085003.}
\lref\Hawkone{S.W. Hawking and G.F.R. Ellis, {\it The Large Scale Structure of
Spacetime}, Cambridge University Press, 1973.}
\lref\Hawktwo{S.W. Hawking, Phys. Rev. D46 (1992) 603.}
\lref\Tip{F.J. Tipler, Phys. Rev. Lett. 37 (1976) 879; Ann. Phys. 108 (1977) 1. }
\lref\Cho{H.T. Cho, Phys. Rev. D56 (1997) 6416.}

\lref\Godel{K. G\"odel, Rev. Mod. Phys. 21 (1949) 447.}
\lref\Thorn{M.S. Morris, K.S. Thorne and U. Yurtsever, Phys. Rev. Lett. 61
(1988) 1446.}
\lref\CSone{A. Vilenkin, Phys. Rev. D23 (1981) 852.}
\lref\CStwo{J.R. Gott, Astrophys. J. 288 (1985) 422.}
\lref\Gott{J.R. Gott, Phys. Rev. Lett. 66 (1991) 1126. }
\lref\DJtHone{S. Deser, R. Jackiw and G. 't Hooft, Ann. Phys. (N.Y.) 
152 (1984) 220. }
\lref\DJtHtwo{S. Deser, R. Jackiw and G. 't Hooft, Phys. Rev. Lett. 68
(1992) 267. }
\lref\DJ{S. Deser and R. Jackiw, Comments Nucl. Part. Phys. 20 (1992) 337.}
\lref\CFG{S.M. Carroll, E. Farhi and A.H. Guth, Phys. Rev. Lett. 68 (1992) 263.}
\lref\CFGO{S.M. Carroll, E. Farhi, A.H. Guth and K.D. Olum, Phys. Rev. D50 (1994) 6190.}
\lref\tH{G. 't Hooft, Class. Quantum Grav. 9 (1992) 1335; 10 (1993) 1023.}
\lref\Gotttwo{M.P. Headrick and J.R. Gott, Phys. Rev. D50 (1994) 7244. }

\lref\AS{P.C. Aichelburg and R.U. Sexl, Gen. Rel. Grav. 2 (1971) 303. }
\lref\DtHone{T. Dray and G. 't Hooft, Nucl. Phys. B253 (1985) 173. }
\lref\DtHtwo{T. Dray and G. 't Hooft, Class. Quant. Grav. 3 (1986) 825. }
\lref\Rob{I. Robinson and A. Trautmann, Proc. R. Soc. A265 (1962) 463.}
\lref\FPV{V. Ferrari, P. Pendenza and G. Veneziano, Gen. Rel. Grav. 20 (1988) 1185. }
\lref\Pen{R. Penrose, unpublished (1974). }
\lref\DE{P.D. D'Eath, {\it Black Holes: Gravitational Interactions}, 
Oxford University Press, 1996. }
\lref\Yurt{U. Yurtsever, Phys. Rev. D38 (1988) 1731.}
\lref\DEP{P.D. D'Eath and P.N. Payne, Phys. Rev. D46 (1992) 658, 675, 694. }
\lref\EGid{D.M. Eardley and S.B. Giddings, Phys. Rev. D66 (2002) 044011. }
\lref\KVen{E. Kohlprath and G. Veneziano, JHEP 0206 (2002) 057. }
\lref\GT{S.B. Giddings and S. Thomas,  Phys. Rev. D65 (2002) 056010. }
\lref\DimL{S. Dimopoulos and G. Landsberg, Phys. Rev. Lett 87 (2001) 161602.}
\lref\Gid{S.B. Giddings, hep-th/0205027.}

\lref\Dmetric{I.T. Drummond, Phys. Rev. D63 (2001) 043503. }


{\nopagenumbers
\rightline{SWAT 02/348}
\vskip1cm
\centerline{\bigbf Constructing Time Machines \footnote{${}^*$}{
\it Invited review article prepared for Int.~J.~Mod.~Phys.~A}}

\vskip1cm

\centerline {\bf G.M. Shore}

\vskip0.5cm
\centerline{\it Department of Physics}
\centerline{\it University of Wales, Swansea}
\centerline{\it Singleton Park}
\centerline{\it Swansea, SA2 8PP, U.K.}

\vskip1cm

{
\parindent 1.5cm{

{\narrower\smallskip\parindent 0pt

The existence of time machines, understood as spacetime constructions
exhibiting physically realised closed timelike curves (CTCs), would raise 
fundamental problems with causality and challenge our current understanding 
of classical and quantum theories of gravity. In this paper, we investigate 
three proposals for time machines which share some common features: 
cosmic strings in relative motion, where the conical spacetime 
appears to allow CTCs; colliding gravitational shock waves, which in 
Aichelburg-Sexl coordinates imply discontinuous geodesics; and the
superluminal propagation of light in gravitational radiation metrics
in a modified electrodynamics featuring violations of the strong equivalence
principle. While we show that ultimately none of these constructions 
creates a working time machine, their study illustrates the subtle 
levels at which causal self-consistency imposes itself, and we consider what 
intuition can be drawn from these examples for future theories.  

\narrower}}}

\vskip2cm

\leftline{SWAT 02/348} 
\leftline{October 2002}

\vfill\eject}

\pageno=1

\newsec{Introduction}

New insight into fundamental theories may often be achieved by studying their 
behaviour in extreme or near-paradoxical regimes. Part of the enduring fascination
of time machines is that they force us to confront basic questions and
assumptions about spacetime inherent in the structure of classical and 
quantum theories of gravity.

In this article, we review three attempts to construct a time machine, which we
understand as a spacetime scenario in which there exist closed signal
trajectories which return to the original spacetime point from which they were 
emitted. Essentially this means showing the existence of physically
realisable closed timelike (or lightlike) curves, which are abbreviated as
CTCs (CLCs). These three examples share certain common features which we feel
makes it interesting to present them together, the paper being a mixture of
review, original work and critique. While none of these constructions
ultimately leads to a working time machine, their analysis involves many quite
subtle issues and reveals how intricate the mechanisms can be by which
quantum theory and general relativity respect causality. 

Time machines can be studied from at least two points of view. One is to try
to formulate general theorems on the occurrence of CTCs in 
exact solutions of Einstein's field equations subject to various axioms,
physical restrictions on the energy-momentum tensor, or realistic
boundary conditions. Such theorems are valuable in making precise the
conditions under which CTCs occur, what exactly constitutes a physically
realisable CTC, and whether their occurrence can be tolerated in 
consistent quantum field theories involving gravity. Examples of this
approach include Hawking's chronology protection conjecture \refs{\Hawktwo}.
See also refs.\refs{\Tip,\Hawkone} for a selection of the relevant literature.

A second approach, which we pursue here, is to try to construct relatively simple
examples of spacetime scenarios with the potential to exhibit causality-violating 
trajectories. This relatively low-tech strategy complements the
axiomatic, theorem-based approach in important ways. Even simple time
machine models can help to sharpen what we should understand by the requirement
that CTCs should be `physically realisable' and what should be considered as
`realistic' boundary conditions. They can uncover essential ingredients for the 
axioms of a general theorem, notably in relation to global rather than local
properties of spacetime. They can be used as probes to uncover an interesting
physical effect -- by imposing the absence of CTCs as a self-consistency
constraint on the kinematics we may deduce, for example, that an event horizon
must form and learn about the conditions under which it does.
And finally, of course, they may succeed!

Simple time machine models typically use one of two fundamental mechanisms
to generate back-in-time motion. One involves rotation -- in some way, the
metric term mixing $t$ and $\phi$ in a rotating spacetime is contrived to
make the periodic angle variable into a timelike coordinate. The canonical 
example of this type of construction is the G\"odel universe \refs{\Godel}.
The second mechanism, which all of our examples exploit, involves propagation that
is effectively superluminal. Superluminal motion, as is well-known in special
relativity, becomes back-in-time when viewed from a certain class of reference
frames, so has at least the potential to lead to causality violation. 
Here, we consider not only superluminal propagation itself but also two
types of spacetime which, like the well-known example of wormholes \refs{\Thorn},
in a sense allow superluminal paths -- the conical spacetime around a cosmic 
string \refs{\CSone,\CStwo}, which can be viewed as flat spacetime with
a missing wedge, and a gravitational shock wave spacetime \refs{\AS}, which 
permits discontinuous geodesics jumping back in time.

We begin in section 2 by recalling some very simple elements of
superluminal motion in special relativity and establish some principles which
we will refer back to at various points in our discussion of the three
time machines. Then we consider probably the most well-known example of a time 
machine construction of this type -- the demonstration by Gott \refs{\Gott}
of the existence of CTCs in the spacetime generated by two cosmic
strings in relative motion. This exploits the fact that a trajectory crossing
the missing wedge in a single cosmic string spacetime can be effectively
superluminal in the sense that it can arrive before a light signal following
a direct path. As Gott then showed, by combining two cosmic strings moving
with velocities exceeding a certain critical value, CTCs exist which encircle
the two strings. The subsequent discussion of whether such CTCs are
physical and consistent with realistic boundary conditions is reviewed 
in section 3. This has proved remarkably fruitful, particularly in terms
of emphasising the importance of global conditions in general theorems
on the occurrence of physical CTCs.

We then try to repeat the success of this idea in a somewhat similar
context. The spacetime generated by a gravitational shock wave is, like the
string spacetime, almost everywhere flat and can be thought of as essentially
two sections of Minkowski spacetime patched together but with an impact parameter 
dependent shift \refs{\DtHone}. Geodesics which cross the spacetime are discontinuous
and jump back in time. This gives another mechanism, analogous to the
wedge-crossing trajectories in the cosmic string spacetime, for effectively
superluminal trajectories. To exploit this to find trajectories which 
return to the past light cone of the emitter (thereby yielding CTCs)
we again need to invoke a second shock wave travelling in
the opposite direction to the first. In the approximation of neglecting
the interaction between the shock waves, we indeed find that CTCs exist 
in which a test particle is successively struck by the two shocks.  
In the Gott cosmic string scenario, the condition for the existence of CTCs
was shown to be violated when the `interaction' between the strings
(in the sense of the conditions of global self-consistency of the two-string 
solution, subject to certain boundary conditions) is included.
In the same spirit, our hope in formulating a shock wave time machine
was that, while the CTCs would not exist in the presence of 
interactions, the precise reasons would shed new light on the kinematics
of the shock wave collision, in particular the possible emergence of an
event horizon in the collision zone. The extent to which this picture is 
realised is described in section 4.

Our final example is rather different in that it involves genuinely
superluminal propagation. As explained in the next section, if we are prepared
to relax the strong equivalence principle, then the action for electrodynamics
in a gravitational field may contain explicit curvature-dependent photon-gravity
interactions. (This SEP-violating effective action is in fact generated by 
vacuum polarisation in QED in curved spacetime, as first shown by Drummond
and Hathrell \refs{\DH}. See also \refs{\Sfive,\Ssix}.)
In sections 2 and 5, we discuss whether superluminal propagation of this
(non-tachyonic) type in general relativity necessarily leads to the same
inevitable causality violations that accompany superluminal propagation
in special relativity. To sharpen this discussion, Dolgov and Novikov \refs{\DN} 
have proposed a construction, again involving two gravitational sources in 
relative motion, in which Drummond-Hathrell superluminal propagation is claimed 
to allow signals to be sent into the past light cone of the emitter. In
section 5, we analyse the Dolgov-Novikov proposal in the context of superluminal
propagation in gravitational radiation spacetimes of Bondi-Sachs type
\refs{\Sfour}. We conclude once more that causality-violating trajectories,
although apparently occurring when the two gravitational sources are
assumed to act entirely independently, do not arise when the full spacetime
involving both sources is considered consistently.

Finally, in section 6 we summarise our conclusions and discuss possible future 
developments.

\vskip2cm

\newsec{Superluminal Propagation and Causality in Special and General Relativity}

It is generally understood that superluminal propagation in special relativity
leads to unacceptable violations of causality. Indeed the absence of tachyons
is traditionally employed as a constraint on fundamental theories. In this
section, we re-examine some basic features of superluminal propagation in 
order to sharpen these ideas in preparation for our subsequent discussion
of the three time machines, especially the Dolgov-Novikov proposal \refs{\DN}
in section 5.

The first important observation is that given a superluminal signal we can
always find a reference frame in which it is travelling backwards in time.
This is illustrated in Fig.~1. Suppose we send a signal from O to A at 
speed $v>1$ (in $c=1$ units) in frame ${\cal S}$ with coordinates $(t,x)$.
In a frame ${\cal S}'$ moving with respect to ${\cal S}$ with velocity 
$u>{1\over v}$, the signal travels backwards in $t'$ time, as follows
immediately from the Lorentz transformation.\foot{From 
the Lorentz transformations,
we have $t'_A = \c(u)t_A (1-uv)$ and $x'_A = \c(u)x_A (1-{u\over v})$
For the situation realised in Fig.~1, we require both $x'_A >0$ and $t'_A <0$,
that is ${1\over v} < u < v$, 
which admits a solution only if $v>1$.}

\vskip0.3cm
{\epsfxsize=11cm\epsfbox{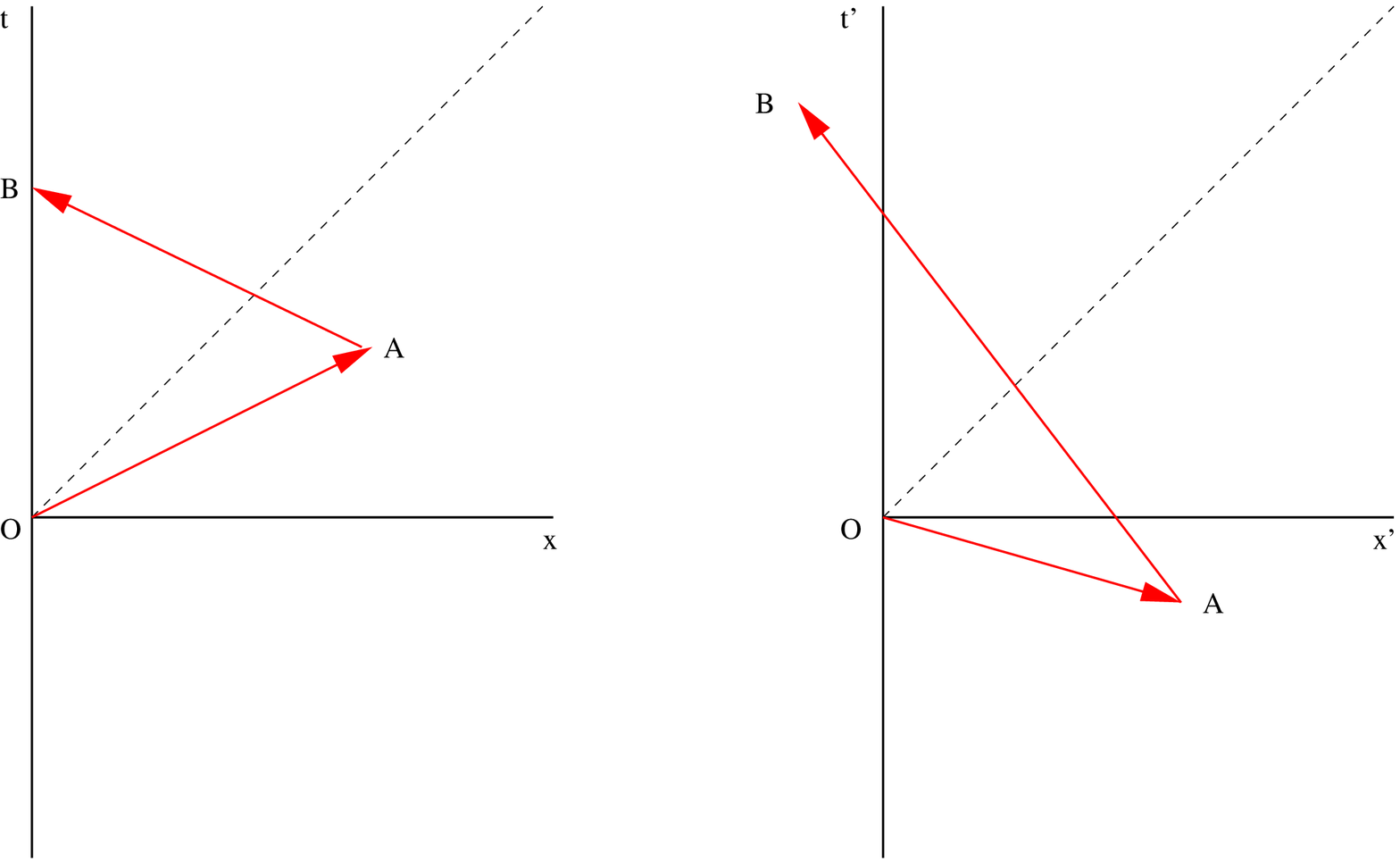}} 
\vskip0.2cm
\noindent{\eightpoint Fig.~1~~A superluminal $(v>1)$ signal OA which is forwards 
in time in frame ${\cal S}$ is backwards in time in a frame ${\cal S}'$ moving
relative to ${\cal S}$ with speed $u>{1\over v}$. However, the return path with
the same speed in ${\cal S}$ arrives at B in the future light cone of O, 
independent of the frame.}
\vskip0.2cm
\noindent The important point for our considerations is that this {\it by itself} 
does not necessarily imply a violation of causality. For this, we require that 
the signal can be returned from A to a point in the past light cone of O. 
However, if we return the signal from A to B with the same speed in frame ${\cal S}$,
then of course it arrives at B in the future cone of O. The situation is physically
equivalent in the Lorentz boosted frame ${\cal S}'$ -- the return signal travels 
forward in $t'$ time and arrives at B in the future cone of O. This, unlike
the assignment of spacetime coordinates, is a frame-independent statement.

The problem with causality arises from the scenario illustrated in Fig.~2.
\vskip0.6cm
{\epsfxsize=5cm\epsfbox{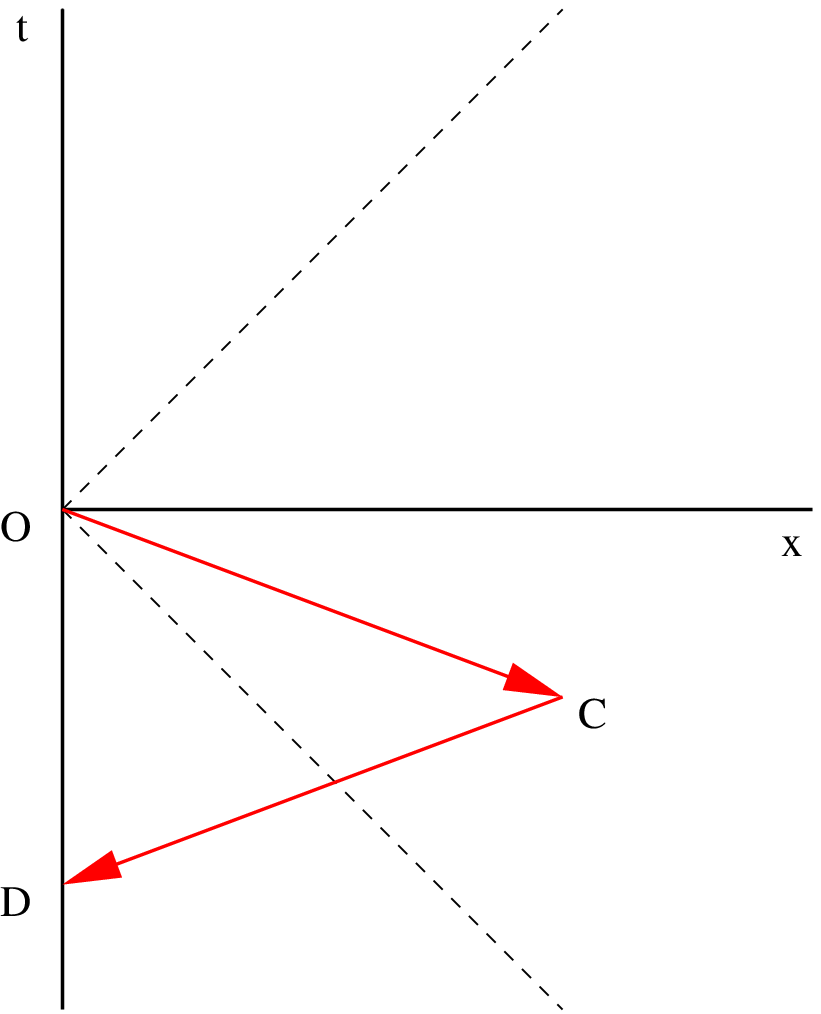}
\vskip0.2cm
\noindent{\eightpoint Fig.~2~~A superluminal $(v>1)$ signal OC which is backwards 
in time in frame ${\cal S}$ is returned at the same speed to point D in the past light
cone of O, creating a closed time loop.}
\vskip0.2cm
\noindent Clearly, if a backwards-in-time signal OC is possible in frame ${\cal S}$,
then a return signal sent with the same speed will arrive at D in the past light
cone of O creating a closed time loop OCDO.
The crucial point is that in special relativity, local Lorentz invariance of the laws 
of motion implies that if a superluminal signal such as OA is possible, then so is one
of type OC, since it is just given by an appropriate Lorentz boost (as in Fig.~1).
The existence of {\it global} inertial frames then guarantees the existence of
the return signal CD (in contrast to the situation in Fig.~1 viewed in the
${\cal S}'$ frame). 

The moral is that {\it both} conditions must be met in order to guarantee the 
occurrence of unacceptable closed time loops -- the existence of a superluminal signal
{\it and} global Lorentz invariance. Of course, since global Lorentz invariance
(the existence of global inertial frames) is the essential part of the structure
of special relativity, we recover the conventional wisdom that in this theory, 
superluminal propagation is indeed in conflict with causality. 

The reason for labouring this elementary point is to emphasise that the situation
is crucially different in general relativity. The {\it weak} equivalence principle, 
which we understand as the statement that a {\it local} inertial (or freely-falling) 
frame exists at each point in spacetime, implies that general relativity is
formulated on a Riemannian manifold. However, local Lorentz invariance is not
sufficient to establish the link discussed above between superluminal propagation
and causality violation. This is usually established by adding to the essential
structure of general relativity a second dynamical assumption. The {\it strong}
equivalence principle (SEP) states that the laws of physics should be identical in
the local frames at different points in spacetime, and that they should
reduce to their special relativistic forms at the origin of each local frame.
It is the SEP which takes over the role of the existence of global inertial frames
in special relativity in establishing the incompatibility of superluminal
propagation and causality.

However, unlike the weak equivalence principle which underpins the essential
structure of general relativity, the SEP appears to be merely a simplifying
assumption about the dynamics of matter coupled to gravitational fields.
Mathematically, it requires that matter or electromagnetism is {\it minimally
coupled} to gravity, i.e.~with interactions depending only on the connections
but not the local curvature. This ensures that at the origin of a local frame,
where the connections may be Lorentz transformed locally to zero, the dynamical
equations recover their special relativistic form. In particular, the SEP would
be violated by interactions which explicitly involve the curvature, such as
terms in the Lagrangian for electromagnetism of the form 
\eqn\sectba{
\C \sim {1\over m^2}~\int dx \sqrt{-g}\biggl[
a R F_{\m\n}F^{\m\n} + b R_{\m\n} F^{\m\l} F^\n{}_\l
+ c R_{\m\n\l\r} F^{\m\n} F^{\l\r} \biggr]
}
for some SEP and conformal breaking mass scale $m$. Such interactions are of more
than speculative interest, since they have been shown by Drummond and
Hathrell \refs{\DH} to arise 
automatically due to quantum effects (vacuum polarisation) in quantum electrodynamics.
They modify the propagation of photons, which no longer follow null geodesics.
Remarkably, such SEP-violating interactions can imply superluminal propagation, 
with photon speeds depending on the local spacetime curvature \refs{\DH}. 

The question of whether this specific realisation of superluminal propagation is
in conflict with causality is the subject of section 5. Notice though that by
violating the SEP, we have evaded the necessary association of superluminal
motion with causality violation that held in special relativity. Referring back to
the figures, what is established is the existence of a signal of type OA,
which as we saw, does not by itself imply problems with causality even though frames 
such as ${\cal S}'$ exist {\it locally} with respect to which motion is backwards in 
time. However, since the SEP is broken, even if a local frame exists in which the 
signal looks like OC, it does {\it not} follow that a return path CD is allowed. 
The signal propagation is fixed, determined locally by the spacetime curvature.

We return to these questions in section 5, where we discuss attempts to use
Drummond-Hathrell superluminal propagation to build a time machine. First, we 
consider two different proposals for time machines which stay within the 
conventional formulation of general relativity but invoke spacetimes with 
special and surprising properties which, in some sense, allow superluminal 
paths.

\vfill\eject

\newsec{A Time Machine from Cosmic Strings?}

The first of our time machines is the well-known construction due to Gott \refs{\Gott},
who showed the existence of CTCs in spacetimes with two cosmic strings in relative
motion. This construction exploits the feature that the spacetime around a cosmic sring
is conical, i.e.~flat everywhere except for a missing wedge where points on the 
opposite edges are identified. This gives rise to the possibility of `short-cut' 
trajectories which can mimic the properties of superluminal paths.

The exterior metric for an infinitely long cosmic string in the $z$ direction is
\eqn\sectca{
ds^2 ~~=~~ - dt^2 + dr^2 + (1-4\m)^2 r^2 d\phi^2 + dz^2
}
where $\m$ is the mass/unit length of the string. (Throughout this paper we use units
with $G=c=\hbar=1$.) To see the conical nature of the spacetime, define
$\hat\phi = (1-4\m)\phi$ so that the metric becomes
\eqn\sectcb{
ds^2 ~~=~~-dt^2 + dr^2 + r^2 d\phi^2 + dz^2
}
with $0\le \phi < (1-4\m)2\pi$. This is clearly Minkowski spacetime apart from a
missing wedge with deficit angle $2\a = 8\pi \m$. The idealisation of an infinitely
long string reduces the problem to one in $(2+1)$ dimensional gravity \refs{\DJtHone}
where there is no dynamics associated with curvature or transverse gravitons and the
physics is determined entirely by global topological features of the spacetime. 

\vskip0.3cm
\noindent{\bf `Non-Interacting' Strings -- A Time Machine}
\vskip0.1cm
Now consider the situation illustrated in Fig.~3.
\vskip0.3cm
{\epsfxsize=7cm\epsfbox{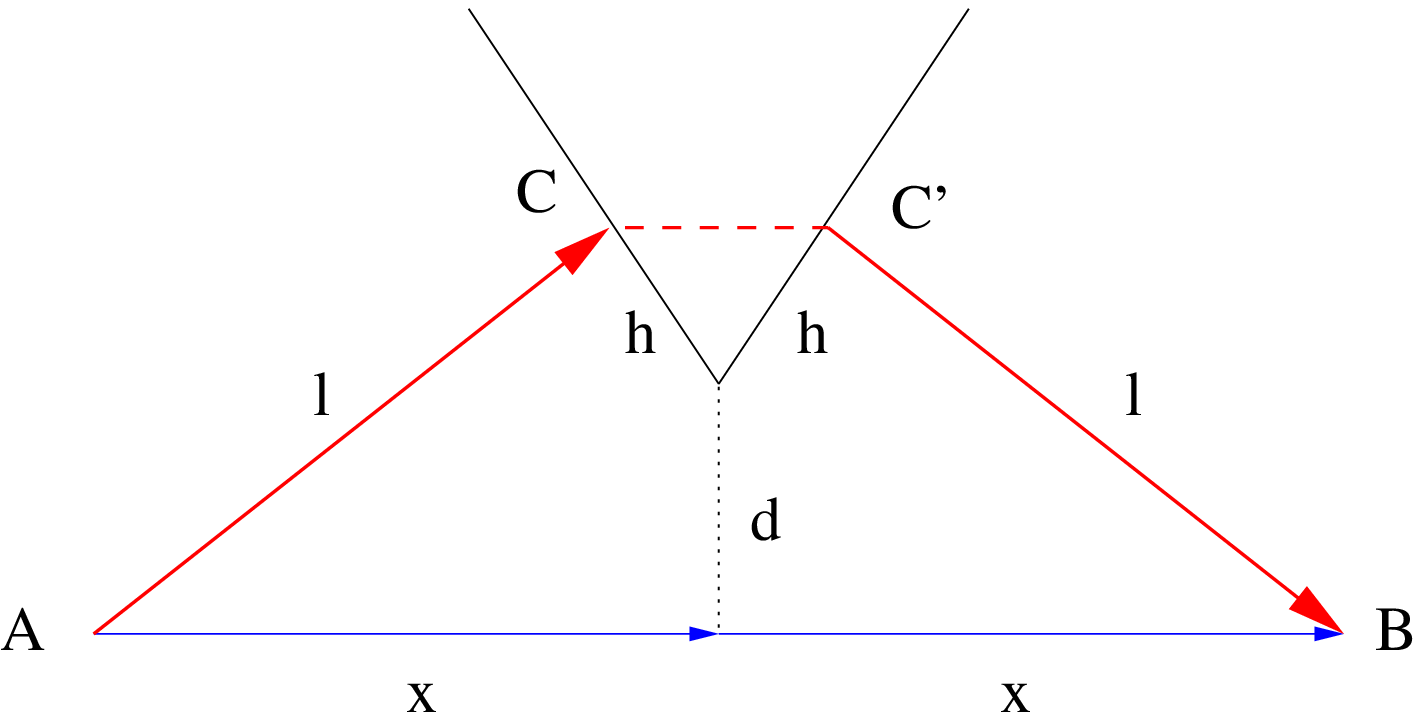}
\vskip0.2cm
\noindent{\eightpoint Fig.~3~~Sketch of the $(x,y)$ plane of the cosmic string
spacetime where the string runs in the $z$ direction from the vertex of the 
missing wedge. The deficit angle of the wedge is $2\a$. Two trajectories are
compared: the direct path AB and the path ACC$^{\prime}$B crossing the wedge, 
where the points C and C$^{\prime}$ are identified. 
The paths intersect the wedge at C and C$^{\prime}$ at right angles.}
\vskip0.2cm
We wish to compare two trajectories. The first is a direct path from A to B; the second
also goes from A to B but via the missing wedge. This is shown as AC followed by 
C$^{\prime}$B
in the figure, where we identify the points C and C$^{\prime}$. The idea is that for some 
choice of parameters, it should be possible to send a signal along the trajectory ACC'B 
so that it takes less time than a light signal sent directly from A to B. This would
effectively realise superluminal propagation, which we can then try to exploit
according to the discussion in the last section to produce back-in-time motion.

From Fig.~3, elementary trigonometry gives the relations
\eqnn\sectcc
$$\eqalignno{
h &= x \sin \a - d \cos \a \cr
x^2 - \ell^2 &= h^2 - d^2 \cr
{}&{}& \sectcc \cr }
$$
from which we can show 
\eqn\sectcd{
1-{\ell^2\over x^2} ~~=~~ \sin^2\a \Bigl(1 - 2{d\over x} \cot\a - {d^2\over x^2}\Bigr)
}

The `effective speed' of a light signal sent along ACC$^{\prime}$B is $v = {x\over \ell}$.
It arrives before a direct light signal along AB if $\ell < x$, i.e.~if the
effective speed $v > 1$. As in the last section, this implies that the ACC$^{\prime}$B signal 
arrives at B having travelled from A {\it back in time} as measured in a frame
moving with velocity $u> {1\over v}$.

This establishes that back-in-time paths across the wedge exist in a frame moving 
relative to the cosmic string with velocity $u > {\ell\over x}$. From eq.\sectcd, this
condition is simply 
\eqn\sectce{
u ~~>~~\cos\a
}
The critical speed is therefore directly related to the deficit angle and hence
the mass/unit length of the string.

So far, we have realised the analogue of the motion OC in Fig.~2. As explained there, 
in order to construct a genuine time machine trajectory we also need to show
the existence of a further back-in-time path CD into the past light cone of the emitter.
The crucial idea in the Gott scenario \refs{\Gott} is that this can indeed be realised
if we have {\it two} cosmic strings moving in opposite directions, each with velocity $u$,
as shown in Fig.~4.

\vskip0.3cm
{\epsfxsize=7cm\epsfbox{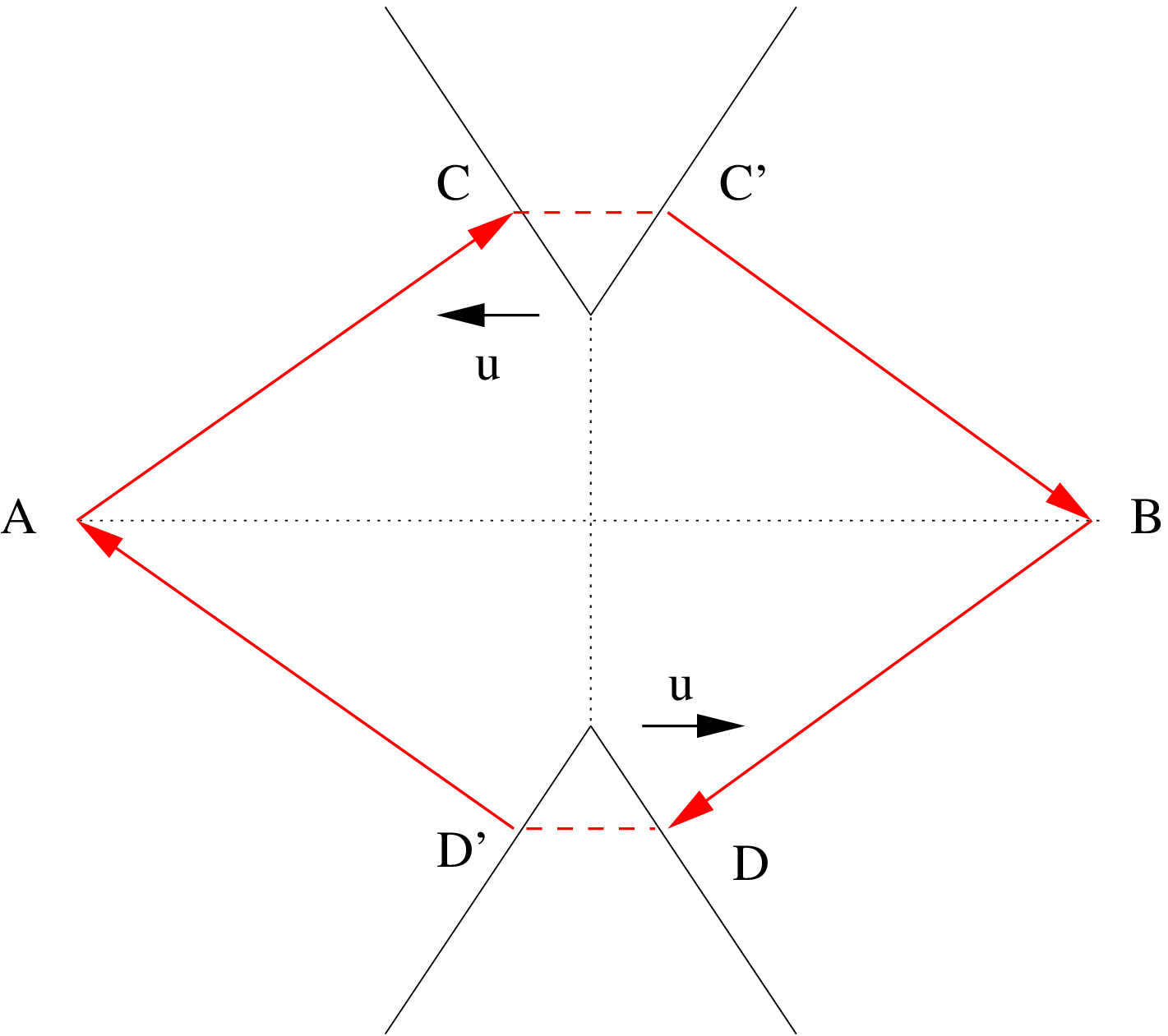}
\vskip0.2cm
\noindent{\eightpoint Fig.~4~~The Gott time machine trajectory. With two cosmic strings
with equal and opposite velocities $u > \cos\a$, the path 
ACC$^{\prime}$BDD$^{\prime}$A can be a CTC. }
\vskip0.2cm
At this point, we assume that the joining of the two cosmic string spacetimes
is trivial (what we may call `non-interacting' strings) and we may freely glue 
together the two patches of Minkowski spacetime. Later we will need to re-examine
carefully the topology of combining the spacetimes in this way. If so, then the same 
arguments that led to the path ACC$^{\prime}$B being effectively superluminal and therefore 
back-in-time apply equally to a return path via the oppositely moving cosmic string
along the wedge path BDD${}^{\prime}$A. The combined path allows a light signal to be sent 
from A back into its own past light cone provided the condition \sectce~is
realised. 

It follows immediately that there exist CTCs looping around the two cosmic strings
provided their velocities satisfy $u >\cos\a$. This is the Gott time machine 
\refs{\Gott}.

\vskip0.3cm
\noindent{\bf `Interacting' Strings and Holonomy}
\vskip0.1cm
Although {\it locally} things seem to be straightforward, patching together single 
cosmic string solutions (or particles in (2+1) dim gravity)
of Einstein's equations to produce new solutions is subject to subtle
and restrictive constraints arising from {\it global} considerations.
For example, although for small mass parameters the combination rule for static
srings is simply additive, if we try to combine two masses $\m_1$ and $\m_2$ where the
corresponding deficit angles satisfy $\a_1 + \a_2 > 2\pi$ then we encounter a global
constraint. The total mass $M$ may exceed ${1\over 2}$ (corresponding to a deficit angle
of $2\pi$) but then the 2-space must be closed and $M$ must equal precisely 1, determined
by the Euler number of the space \refs{\DJtHone}.

Patching together cosmic string spacetimes when the strings are in motion, even for
small masses, also raises interesting global considerations. This can be viewed in terms 
of holonomy. In a curved spacetime, parallel transport of a vector in the tangent space at 
given point around a closed curve results in a transformation of that vector due to
the non-vanishing curvature tensor. The group of these transformations is the holonomy group,
which for a general (pseudo) Riemannian space is just the Lorentz group SO(3,1). 

In the (2+1) dim flat spacetime corresponding to an infinite cosmic string, parallel 
transport around a closed curve encircling the string also produces a non-vanishing
SO(2,1) holonomy transformation due to the deficit angle. For a stationary string, 
this is just a rotation equal to the deficit angle. The group element may be written
in the form
\eqn\sectcf{
T ~~=~~ \exp\bigl( 8\pi i P^\m \S_\m\bigr)
}
where $\S_\m$ are the three SO(2,1) generators and $P^\m$ is the string energy-momentum
vector. For a stationary string, $P^0 = \m$, $P^1=P^2 = 0$, $\S_0 =
\left(\matrix{0&0&0\cr
0&0&-i\cr
0&i&0\cr}\right)$
and $T$ is the rotation matrix through an angle $2\a = 8\pi \m$ in the $(x,y)$ plane,
$R_\a = 
\left(\matrix{1&0&0\cr
0&\cos 2\a &\sin 2\a \cr
0&-\sin 2\a &\cos 2\a\cr}\right)$.
A boost with velocity $u$ is represented by $L_u = 
\left(\matrix{\cosh\b &\sinh\b & 0\cr
\sinh\b &\cosh\b & 0 \cr
0 &0 &1 \cr} \right)$
where the rapidity is defined through $\tanh\b = u$.

The element of the holonomy group for a moving cosmic string is thus $L_u^{-1} R_\a L_u$.
So for the two cosmic string Gott scenario, a loop encircling both strings is associated
with the holonomy \refs{\DJtHtwo,\DJ,\CFG}
\eqn\sectcg{
T = L_u R_\a L_u^{-1} ~L_u^{-1} R_\a L_u
}
since the strings are moving in opposite directions. If we now try to identify this with 
the holonomy of a single string with mass/unit length $M$ and deficit angle $A=4\pi M$)
with some velocity $U$, then we would write
\eqn\sectch{
T = L_U^{-1} R_A L_U
}
Taking the trace and equating these two expressions, we find
\eqn\sectci{
{\rm tr}~ R_A ~~=~~ {\rm tr}~ L_u^2 R_\a L_u^{-2} R_\a
}
which gives
\eqn\sectcj{
\cos A ~~=~~ \cos^2\a ~-~ \sin^2\a ~\cosh2\b
}
or 
\eqn\sectck{
\sin{A\over2} ~~=~~ \cosh\b ~\sin\a
}
For non-zero velocities, this gives a non-trivial and non-linear addition rule
for masses, viz.
\eqnn\sectcl
$$\eqalignno{
\sin 2\pi M ~~&=~~ \cosh\b ~ \sin 4\pi\m \cr
{}&=~~{1\over\sqrt{1-v^2}}~\sin4\pi\m \cr
{}&{}& \sectcl \cr}
$$

Now for this identification to work, clearly the r.h.s.~of eq.\sectck ~must be less than 1.
This gives the condition \refs{\DJtHtwo,\DJ,\CFG}
\eqn\sectcm{
\cosh\b ~\sin\a ~~< ~~1
}
or equivalently
\eqn\sectcn{
u ~~< ~~ \cos\a
}
This is exactly the {\it opposite} of the criterion for the existence of CTCs
in the Gott time machine scenario.

From eq.\sectcf, the Gott condition $u > \cos\a$ therefore appears to correspond
to a holonomy element representing a boost rather than a rotation. Equivalently, the 
combined two-string system would have a spacelike energy-momentum vector
($P^2 > 0$) corresponding to tachyonic motion. Although both constituent strings
have subluminal velocities $u < 1$, the non-linear energy-momentum addition
rules following from the holonomies imply that CTCs only arise for velocities
for which the combined system is tachyonic and hence unphysical \refs{\DJtHtwo,\DJ,\CFG}. 

This appears to be a clean resolution of the cosmic string time machine puzzle, showing
neatly that CTCs are necessarily associated with physically unrealisable
energy-momentum conditions. However, even closer analysis shows that the situation
is not quite so clear-cut. If we consider parallel transport of spinors rather than 
just vectors around a closed curve, the relevant holonomy group becomes SU(1,1), the
double cover of SO(2,1). The group manifold of SU(1,1) reduces to that of SO(2,1)
under an identification of points and, as shown explicitly in ref.\refs{\CFGO}, it
is only under this identification that the holonomy element associated with the Gott
string pair can be represented as the exponential of a spacelike generator. In
the full SU(1,1) group, the Gott holonomy element cannot be represented as the  
exponential of a generator. (In terms of the geometry of the group manifold, this
corresponds to a point which cannot be reached from the identity by following a
geodesic path.) In this more general context, it is {\it not} true that the Gott
condition implies a spacelike energy-momentum vector. Indeed, the whole idea that
the energy-momentum vector in (2+1) dim gravity can be defined by its identification
with holonomy elements becomes problematic \refs{\Gotttwo}.

It remains the case, however, that the Gott spacetime contains CTCs at spacelike infinity
\refs{\DJtHtwo}. In ref.\refs{\DJtHtwo}, this is regarded as an unphysical boundary 
condition, but this assertion is challenged with some justification in ref.{\Gotttwo}.
After all, unless we rule out the existence of CTCs as unphysical from the start
and use this as a selection criterion as to what matter content and boundary conditions we
consider to be `physical' or `unphysical', why should their absence be required as
a boundary condition? Presumably, {\it if} we were able to make sense of quantum field theory
on background spacetimes involving CTCs (see e.g.~ref.\refs{\Gotttwo} for a selection of references), 
then we would be happy to regard such spacetimes and boundary conditions as physical. 

Finally, an important further result on the Gott scenario is developed in 
refs.\refs{\CFG,\CFGO,\tH}. In the first two of these papers, it is shown that the Gott
time machine cannot be created in an open universe with a timelike total energy-momentum.
Essentially, in an open universe, there is not enough energy to satisfy the Gott condition.
On the other hand, 't Hooft showed \refs{\tH} that in a closed universe, the spacetime collapses
to zero volume, a big crunch, before the Gott CTCs have time to form. So it seems that
in a universe with timelike total energy-momentum which is initially free from CTCs, 
the Gott mechanism does not in fact create a time machine.

\vfill\eject

\newsec{A Time Machine from Gravitational Shock Waves?}

The spacetime describing a gravitational shock wave exhibits the unusual
property that, depending on their impact parameter, geodesics which cross the shock 
may experience a discontinuous jump backwards in time. In this section, we investigate
whether this phenomenon can be exploited to construct a time machine.

A gravitational shock wave is described by the Aichelburg-Sexl metric \refs{\AS} \foot{
The Aichelburg-Sexl metric can be derived, for example, by taking the Schwarzschild
solution for a mass $M$ and boosting along the $z$ direction with Lorentz parameter
$\c$, taking the infinite boost limit $\c\rightarrow\infty$ with $\c M$ fixed.
An alternative method \refs{\DtHone} involves a technique of `cut-and-paste' along
a null hypersurface.
For a detailed description of the A-S metric and the collision of two shock waves, 
see e.g.~refs.\refs{\DE,\DtHtwo}. The idealisation of an infinitely thin shock
can of course be smoothed out, in which case the geodesics across the shock simply 
change very rapidly rather than actually discontinuously.}
\eqn\sectda{
ds^2 = - du dv + f(r) \d(u) du^2 + dr^2 + r^2 d\phi^2
}
with light-cone coordinates $u=t-z$, $v=t+z$. This describes an  
axisymmetric, plane-fronted shock wave advancing in the 
positive $z$ direction at the speed of light $(u=0)$
with a profile function $f(r)$ in the transverse direction. 
It separates two regions of spacetime which are manifestly Minkowskian in
standard coordinates. The Einstein equation
\eqn\sectdb{
R_{uu} = 8\pi T_{uu}
}
is satisfied for a matter source $T_{uu} = \r(r)\d(u)$, provided the profile function
obeys
\eqn\sectdc{
\D f(r) = - 16\pi \r(r)
}
where $\D$ is the transverse 2 dim Laplacian. Two matter sources are of particular 
interest: a particle with $\r(r)=\m \d({\underline x})$, for which the profile function
is $f(r) = -4\m \ln {r^2\over L^2}$ for some arbitrary scale $L$, and a homogeneous beam
$\r(r) = \r = {\rm const} ~(r < R)$ for which $f(r) = -4\pi \r r^2 ~(r < R)$
and $-4\pi \r R^2 \bigl(1+ \ln {r^2\over R^2}\bigr) ~(r > R)$. The geodesics for an
infinite homogeneous shell were discussed by Dray and 't Hooft \refs{\DtHtwo} and 
generalised to a finite homogeneous beam in arbitrary dimensions by Ferrari, Pendenza 
and Veneziano \refs{\FPV}.
(We always consider impact parameters less than the beam size $R$ in those parts of the
following discussion where we specialise to homogeneous beams.)

Geodesics in this spacetime satisfy the following equations (${}^{.}$ denotes 
differentiation w.r.t.~the affine parameter and $'$ differentiation w.r.t.~$r$):
\eqnn\sectdd
$$\eqalignno{
{}&\ddot{u} = ~~0~~ \cr
{}&\ddot{v} - f(r) \d'(u){\dot u}^2 - 2 f'(r) \d(u) {\dot u}{\dot r} ~~=~~0 \cr
{}&\ddot{r} - {1\over2}f'(r)\d(u) {\dot u}^2 + r{\dot \phi}^2 ~~=~~0 \cr 
{}&\ddot{\phi} + {2\over r} {\dot r}{\dot \phi} ~~=~~0 \cr
{}&{}& \sectdd \cr }
$$
In particular, a null geodesic in the negative $z$ direction making a head-on
collision with impact parameter $b$ with the shock is described by
\eqnn\sectde
$$\eqalignno{
v ~~&=~~f(b) \theta(u) + \tan^2{\varphi\over2} ~u\theta(u) \cr
r~~&=~~b - u \tan{\varphi\over2} \cr
{}&{}& \sectde \cr}
$$
where the deflection angle $\varphi$ satisfies $\tan{\varphi\over2} = -{1\over2}f'(b)$.
Notice that for sufficiently large impact parameters, $\varphi$ can exceed $\pi/2$
and the geodesic is reflected back in the positive $z$ direction.
\vskip0.2cm
{\epsfxsize=6cm\epsfbox{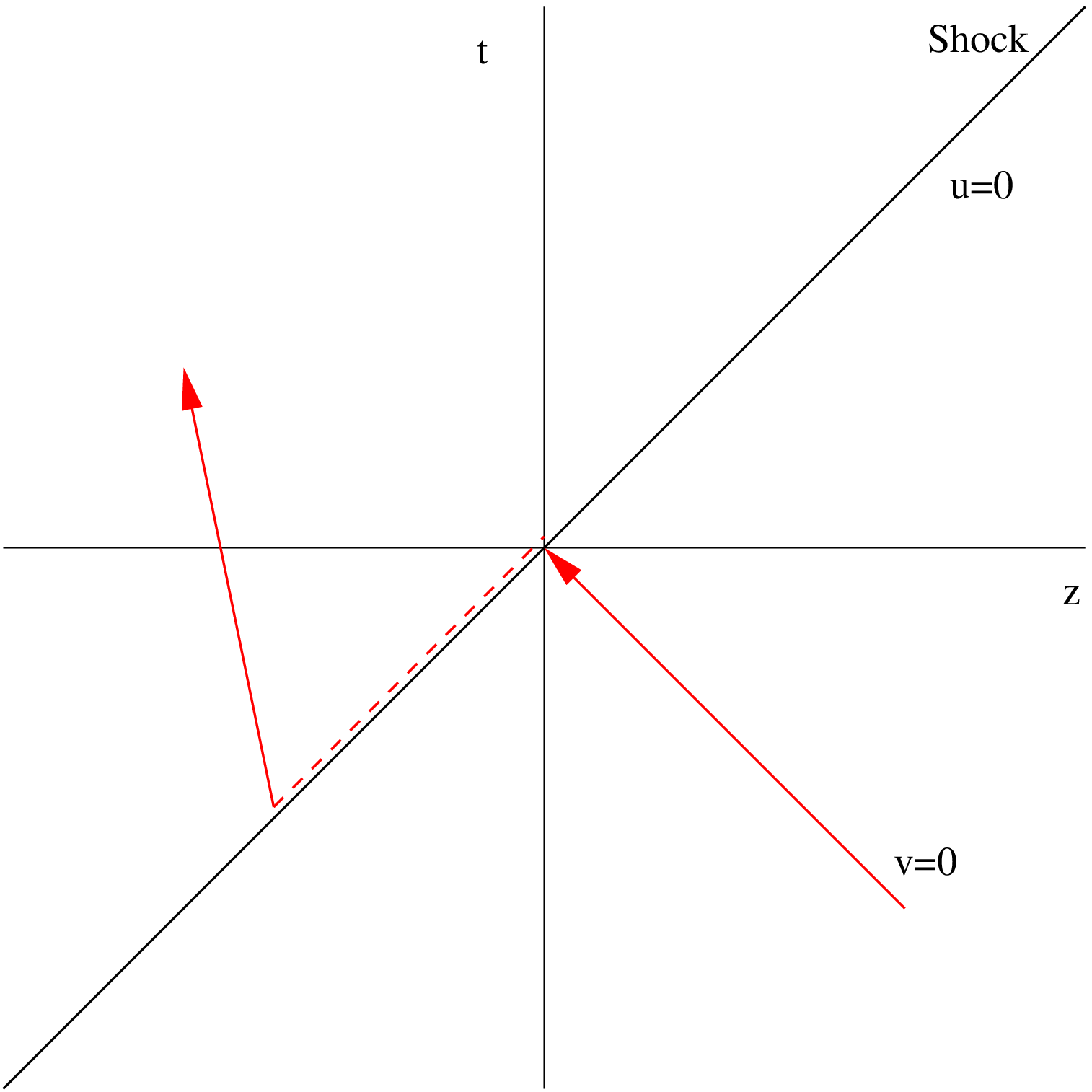}}
\hskip1cm 
{\epsfxsize=6cm\epsfbox{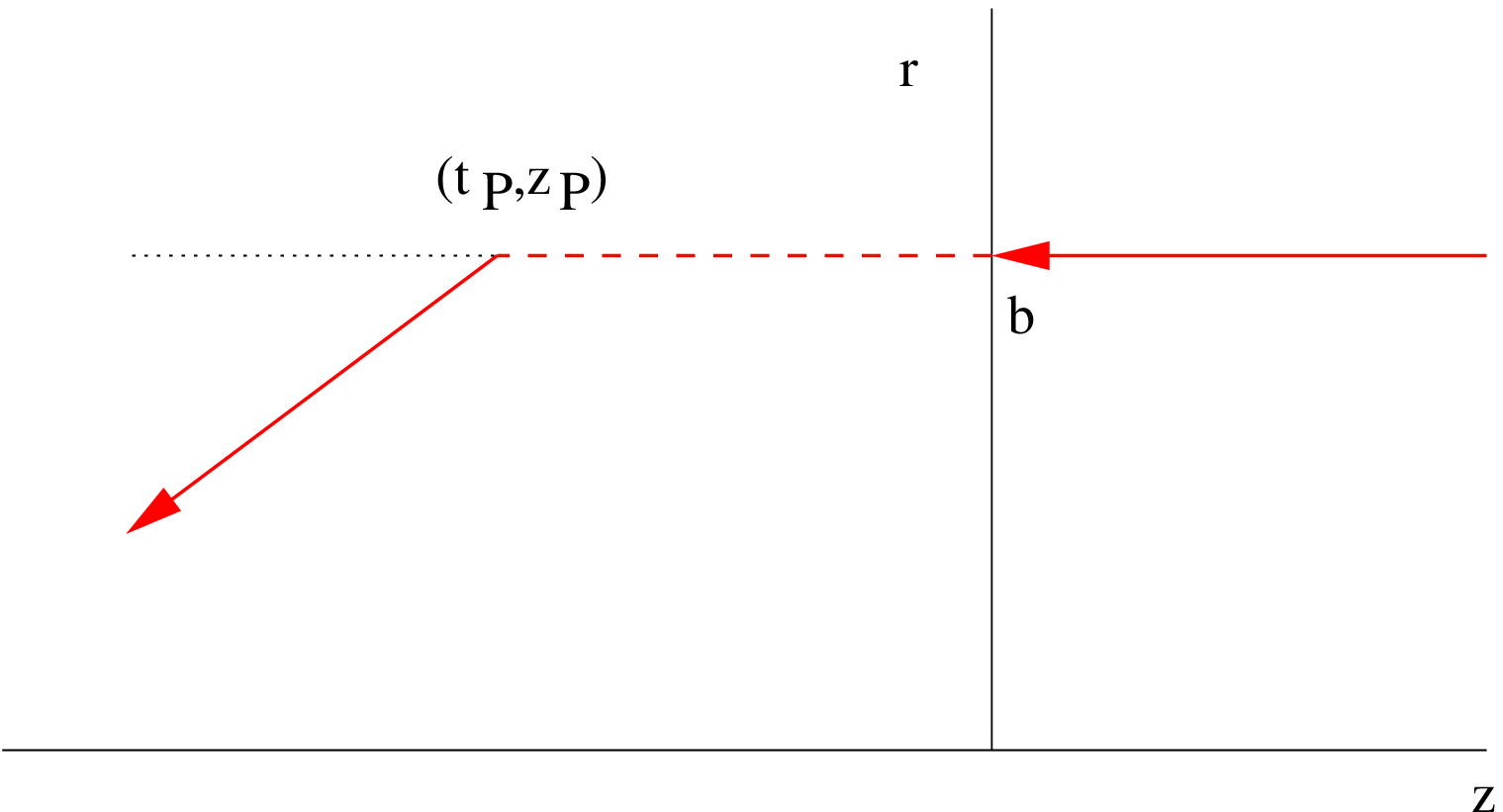}} 
\vskip0.01cm
\noindent{\eightpoint Fig.~5~~Sketches of null geodesics in the Aichelburg-Sexl metric 
in the $(t,z)$ and $(r,z)$ planes. The incoming geodesic experiences a discontinuous jump
on collision with the shock wavefront at  $u=v=0$ from $t=z=0$ to $t_P=z_P={1\over2}f(b)$
(shown as a dotted line in the figures). For $b>L$ (particle) or all $b$ (homogeneous beam),
the shift is backwards in time, $t_P < 0$. The geodesic is also deflected towards the $r=0$ 
axis by an angle $\varphi$ where $\tan{\varphi\over2} = -{1\over2}f'(b)$.}

The important observation is that the geodesics make a discontinuous jump\foot{It should
immediately be noted that this discontinuity can be removed by a discontinuous change of
coordinates in which the geodesics are continuous (though not differentiable) at the shock, 
but where the manifest Minkowski nature of the external metric is lost (see, 
e.g.~ref.\refs{\DE}). For this reason, we prefer to present the
discussion of possible time paradoxes in the original Aichelburg-Sexl form, where 
the use of standard Minkowski coordinates highlights the peculiar effects of the 
shock wave on geodesic paths.} (see Fig.~5)
from the collision surface $u=v=0$ to $u=0$, $v_P = f(b)$, i.e. $t_P=z_P= {1\over2}f(b)$,
which in the case of a homogeneous beam in particular is always negative for all
impact parameters. Collision with the shock front therefore induces a jump {\it backwards 
in time}. This is the effect we will use to try and construct a time machine.

Before this, we also need the formulae for null geodesics making a collision with
the shock wave at a general angle of incidence, say $\psi$. In this case, the geodesic
is described by
\eqnn\sectdf
$$\eqalignno{
v~~&=~~ u \big(\tan{\psi\over2}\bigr)^2 \theta(-u) + 
\Bigl[f(b) + u \big(\tan{\varphi'\over2}\bigr)^2\Bigr] \theta(u) \cr
r~~&=~~ \Bigl[b-u\tan{\varphi\over2}\Bigr] \theta(-u) +
\Bigl[b - u \tan{\varphi'\over2}\Bigr] \theta(u) \cr
{}&{}& \sectdf \cr }
$$
The incident $(\psi)$ and deflected $(\varphi')$ angles satisfy the elegant addition
formula\refs{\FPV}
\eqn\sectdg{
\tan{\varphi'\over2}~~=~~\tan{\psi\over2}+\tan{\varphi\over2}
}
with the definition $\tan{\varphi\over2}= -{1\over2}f'(b)$ as before.

There is a further effect which will be important to us, viz.~geodesic focusing.
From the formulae above, we see that after collision the geodesics will intersect
the $r=0$ axis at $(t_F,z_F)$ where
\eqnn\sectdg
$$\eqalignno{
t_F~~&=~~{1\over2}f(b) - {b\over f'(b)} - {1\over4}f'(b) b \cr
z_F~~&=~~{1\over2}f(b) + {b\over f'(b)} - {1\over4}f'(b) b \cr
{}&{}& \sectdg \cr }
$$
For a particle shock, this results in a caustic, but for the special case of
a homogeneous beam, $f(r) = -\a r^2$ (where from now on we use the notation
$\a \equiv 4\pi \r$ to avoid unnecessary factors of $4\pi$), the geodesics actually 
{\it focus} to the single point $u_F={1\over\a}$, $v_F=0$, i.e.
\eqn\sectdh{
t_F = - z_F = {1\over 2\a}
}
The greater the energy density in the beam, the closer the focal point is to the collision
surface. This is illustrated in the $(t,z)$ and $(r,z)$ plane sketches in Fig.~6 and 
the 3D plot in Fig.~7. For a finite beam ($\r(r) = 0$ for $r>R$) this is simply a focal 
point\refs{\FPV}, whereas for an infinite shell it produces a focusing singularity 
\refs{\DtHtwo}.

\vskip0.2cm
{\epsfxsize=6cm\epsfbox{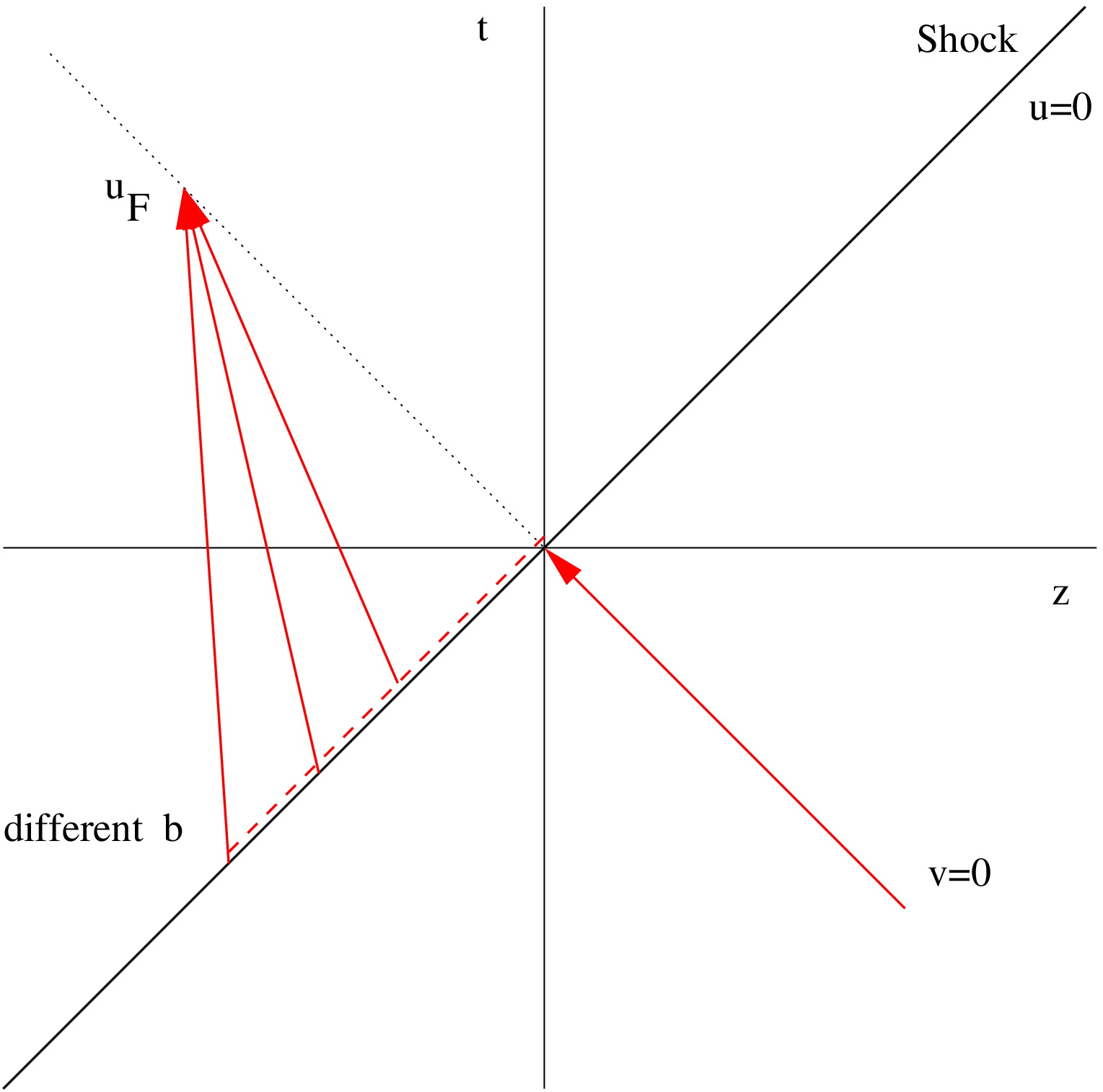}}
\hskip1cm 
{\epsfxsize=6cm\epsfbox{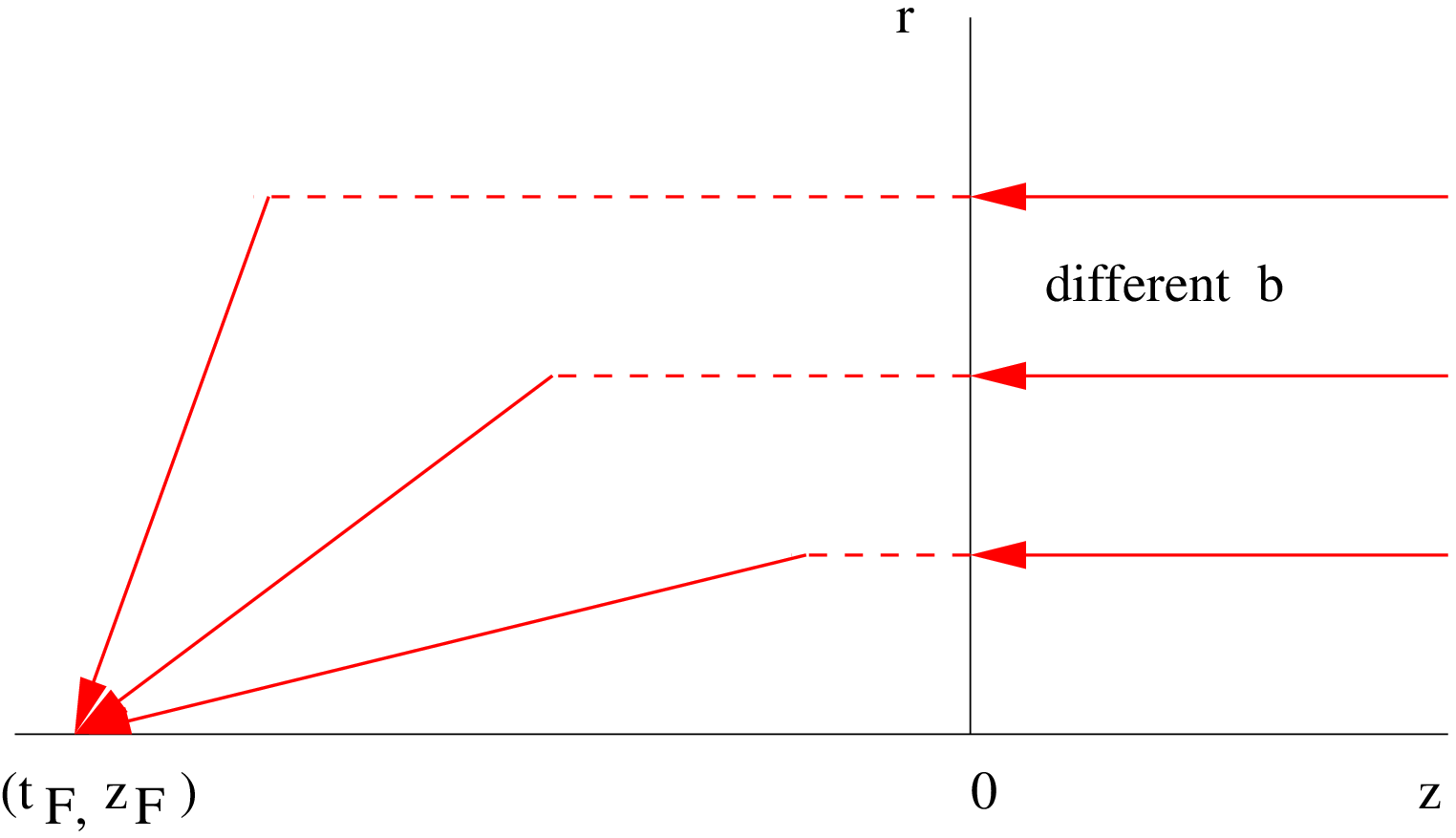}} 
\vskip0.01cm
\noindent{\eightpoint Fig.~6~~Sketches in the $(t,z)$ and $(r,z)$ planes 
showing the focusing of geodesics with different
impact parameters $(b)$ at the point $u_F= {1\over\a}$ following collision with a
homogeneous beam with profile $f(r) = -\a r^2$.}

\vskip0.2cm
{\epsfxsize=7cm\epsfbox{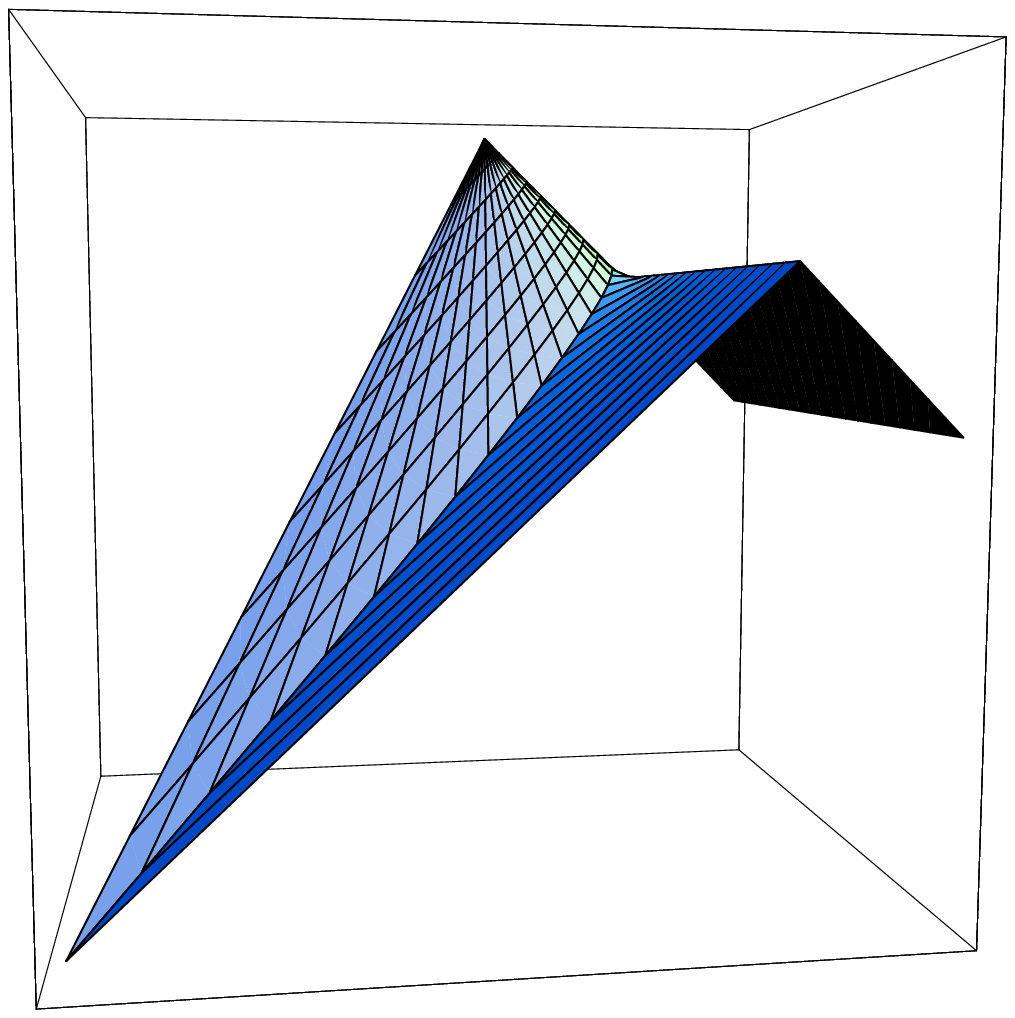}} 
\vskip0.01cm
\noindent{\eightpoint Fig.~7~~3D plot in $(z,r,t)$ coordinates of the paths illustrated in 
Fig.~5. The transverse $r$ coordinate is zero at the back of the diagram with larger
values of the impact parameter $r=b$ illustrated towards the front. The discontinuous
jump of the geodesics followed by deflection towards $r=0$ and focusing is clearly seen.}
\vskip0.1cm

\noindent{\bf Non-Interacting Shocks -- A Time Machine}
\vskip0.1cm
These discontinuous jumps backwards in time therefore appear to provide a promising
mechanism to construct a time machine. However, as we emphasised in section 2, it is
not at all sufficient to find a coordinate system in which motion occurs backwards
in `time'. A genuine time machine requires that a signal can be sent into the past light
cone of the emitter. It is clear that this cannot be achieved by considering the geodesics
of only one shock wave.

Inspired by the Gott proposal with cosmic strings, we therefore consider the spacetime
formed by two gravitational shock waves moving towards each other along the 
$z$ direction with zero impact parameter. We shall initially assume that we may 
neglect the interaction of the shocks themselves, so that in particular
the shock fronts remain unchanged in the region $u,v>0$. (Later, as in the cosmic
string example, we will need to examine the effects of the interaction of the two
gravitational sources more closely.) For simplicity, we 
choose the shocks to be generated by finite homogeneous beams, with profile functions
$f_1(r) = -\a r^2$ and $f_2(r) = -\b r^2$  (for $r < R$) respectively.

We therefore consider the following scenario, illustrated in Fig.~8.
\vskip0.2cm
{\epsfxsize=8cm\epsfbox{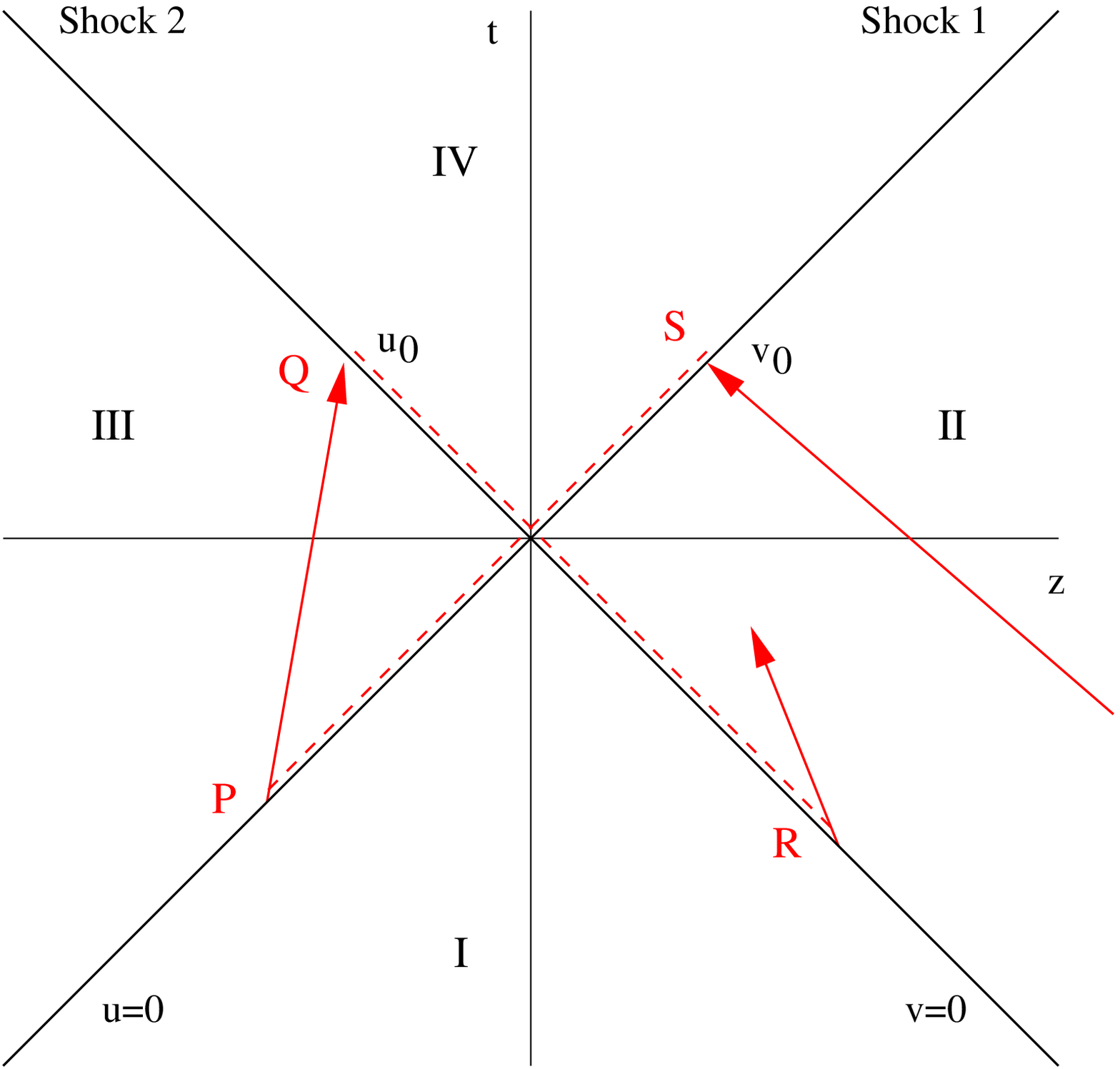}} 
\vskip0.01cm
\noindent{\eightpoint Fig.~8~~Two gravitational shock waves, comprising homogeneous
beams with energy densities given by $4\pi\r(r) = \a r^2$ and $\b r^2$ respectively,
moving in opposite directions along the $z$ axis intersect at $u=v=0$ with zero
impact parameter. They are assumed not to interact so that the shock fronts continue
unchanged along the trajectories $u=0$ or $v=0$.
The `time machine' trajectory SPQR comprises
a null geodesic colliding with shock wave 1 at point S, jumping back in time
to P and reversing its direction, colliding with shock 2 at Q and again jumping
back in time to R. For suitable impact parameters, R lies in the past lightcone of S.}

\noindent Consider a signal following a null geodesic which is struck head-on by shock 1 at 
point $S$ with impact parameter $a$. It experiences a discontinuous jump back in time (and
back in the longitudinal direction $z$) to P. For a suitable choice of impact parameter,
its angle of deflection $\varphi_a$ can be sufficiently large that its direction
along the $z$ axis is actually changed and it is reflected back into the path of
the oncoming second shock wave. It collides with shock 2 at point Q in an oblique
collision with impact parameter $b$ and angle of incidence $\psi$. A further
discontinuous jump back in time takes the signal to R, where it continues its further
deflected path. We wish to show that for suitable parameter choices, the spacetime
interval RS is {\it timelike}, i.e.~R really does, as the figure suggests, lie in
the past light cone of S.

The geodesic equations governing this sequence of events are as follows.  
Approaching S, the initial signal follows the null geodesic
\eqnn\sectdi
$$\eqalignno{
v~~&=~~ v_0 \cr
r~~&=~~ a \cr
{}&{}& \sectdi \cr }
$$
Following the impact with shock 1 and jump to P, the signal along PQ follows
the geodesic
\eqnn\sectdj
$$\eqalignno{
v~~&=~~ v_0 + f_1(a) + u \Bigl(\tan{\varphi_a\over2}\Bigr)^2 \cr
r~~&=~~ a - u \tan{\varphi_a\over2} \cr
{}&{}& \sectdj \cr }
$$
with $\tan{\varphi_a\over2} = -{1\over2}f'_1(a)$. Clearly,
\eqnn\sectdk
$$\eqalignno{
v_P~~&=~~ v_0 + f_1(a) \cr
r_P~~&=~~ a \cr
{}&{}& \sectdk \cr }
$$
Under the no-distortion approximation, the signal then makes an oblique impact
with the second shock front at Q where $v_Q=0, u_Q = u_0, r_Q=b$,
with $v_0$ and $b$ to be determined, and the angle of incidence 
$\psi = \pi - \varphi_a$. From the PQ geodesic equations, we deduce
\eqnn\sectdl
$$\eqalignno{
v_Q~~&=~~ u_0~~=~~ -\bigl(v_0 + f_1(a)\bigr) 
\Bigl(\tan{\varphi_a\over2}\Bigr)^{-2}\cr
r_Q~~&=~~ b~~=~~ a ~+~ \bigl(v_0 + f_1(a)\bigr) 
\Bigl(\tan{\varphi_a\over2}\Bigr)^{-1}   \cr
{}&{}& \sectdl \cr }
$$ 
After the second jump, the trajectory RS is therefore
\eqnn\sectdm
$$\eqalignno{
u ~~&=~~u_0 + f_2(b) + v \Bigl(\tan{\theta\over2}\Bigr)^{2} \cr
r~~&=~~ b - v \tan{\theta\over2} \cr
{}&{}& \sectdm \cr }
$$
where 
\eqnn\sectdn
$$\eqalignno{
\tan{\theta\over2}~~&=~~ \tan{\varphi_b\over2} + \tan{\psi\over2} \cr
{}&=~~\tan{\varphi_b\over2} + \Bigl(\tan{\varphi_a\over2}\Bigr)^{-1} \cr
{}&{}& \sectdn \cr }
$$
In particular, the point R is 
\eqnn\sectdo
$$\eqalignno{
u_R~~&=~~ u_0 + f_2(b) \cr
r_R~~&=~~ b \cr
{}&{}& \sectdo \cr }
$$

The spacetime interval $\D s^2{}_{\rm RS}$ is simply
\eqn\sectdp{
\D s^2{}_{\rm RS} ~~=~~ -(u_S - u_R) (v_S - v_R) + (r_S - r_R)^2
}
since the spacetime is flat in region II, and we find (eliminating $u_0$
and $v_0$ in favour of $f_1(a)$ and $f_2(b)$),
\eqn\sectdq{
\D s^2{}_{\rm RS} ~~=~~ - f_1(a) f_2(b) ~-~(a-b) 
\biggl(f_1(a) \Bigl(\tan{\varphi_a\over2}\Bigr)^{-1} + f_1(b) 
\Bigl(\tan{\varphi_a\over2}\Bigr)\biggr)
}

So far, the discussion has applied to both particles and beams as the source of
the shockwaves as the profile functions $f_1(r)$ and $f_2(r)$ have not been 
specified. For definiteness, we now consider homogeneous beams with profiles
$f_1(r) = -\a r^2$ and $f_2(r) = -\b r^2$. In this case, the condition 
\sectdq~becomes
\eqn\sectdr{
\D s^2{}_{\rm RS} ~~=~~ - (\a \b a b) ~b^2 ~~+~~\Bigl(1-{b\over a}\Bigr)~ a^2 
}
The parameter count is as follows. $\a,\b$ specify the energy densities of the
two beams, so are free parameters for us to choose. Likewise, the initial
signal trajectory, which we may also choose freely, is specified by the two
parameters $a$ (impact parameter) and $v_0$ (location of the initial collision). 
Using eq.~\sectdl, we can trade $v_0$ for the second impact parameter $b$, which 
may then be considered as the other free parameter. So in eq.~\sectdr, all the 
parameters can be chosen freely.

In particular, we can certainly choose the free parameters such that
\eqn\sectds{
\D s^2{}_{\rm RS} ~~<~~ 0 
}
i.e. RS is {\it timelike} with R indeed in the past light cone of S.\foot{
It is perhaps also interesting to note that if we allow the signal to continue
on from R in the negative $z$ direction according to eq.~\sectdm, it will collide
again with shock 1 at a point T with $v_T < v_S$. The null geodesic path
therefore continues to cycle round a similar loop to SPQRT, getting closer to
the origin $u=v=0$ with decreasing $r$ in each loop.}
In other words, the trajectory SPQR describes a time machine! 

\vskip0.5cm
\noindent{\bf Interacting Shocks}
\vskip0.3cm
As explained in the introduction, the motivation for this construction is not so
much the belief that a time machine would actually exist in this spacetime, but
that the reason preventing its existence would be interesting and shed new light
on the physics of gravitational shock wave collisions. Indeed, there is much
current interest in shock wave collisions in connection with the possible production
of black holes in transplanckian energy collisions, especially since certain
large extra dimension theories can accommodate a lowering of the effective
four dimensional Planck energy to the $TeV$ scale with the consequence that
transplanckian scattering could actually be realised in $TeV$ colliders such as the 
LHC (see, e.g.~refs.\refs{\DimL,\GT,\Gid}).

There are three particularly significant effects of the shock wave interaction
that we have yet to consider. First, in the future region of the collision
surface, the spacetime is curved, gravitational radiation may be emitted, and 
in general the metric is difficult to determine even perturbatively.(See however 
ref.\refs{\DtHtwo} for a solution in the case of infinite shells 
in terms of `Robinson's nullicle' \refs{\Rob}.) Second, the shock wavefronts
themselves are distorted and, in the case of homogeneous beams, become
concentric spherical shells converging to focal points. Third, for certain
impact parameters, a closed trapped surface may form in the collision region.

The most interesting of these from our perspective would be the latter. 
The identification of a closed trapped surface in the colliding shock wave geometry
was first made by Penrose \refs{\Pen} (see also \refs{\DE}) and also discussed 
by Yurtsever \refs{\Yurt}.
Recently, this construction has been generalised to higher dimensions and 
non-zero impact parameters by Eardley and Giddings \refs{\EGid}, and to the case of
beams by Kohlprath and Veneziano \refs{\KVen}. Indeed, 
the original, rather speculative, motivation for pursuing this `time machine'
construction was that the impact parameters of the trajectories which would be 
required to produce a causality-violating loop would be related to the
location of a possible trapped surface. In that case, the preservation of 
causality would be linked to the presence of an event horizon. 

Unfortunately, the reason why the time machine fails is rather less exotic and
concerns the first two properties of the interaction. At first sight, the
future curved zone appears to be simply the wedge IV on the 
spacetime diagram Fig.~8, i.e.~the future region of the initial collision
surface $u=v=0$. However, as we see below, the identification of the `collision
surface' and thus the curved zone in Aichelburg-Sexl coordinates is slightly
more subtle. The other key ingredient is the distortion of the shock fronts
themselves.

The shock wavefronts are generated by a congruence of null geodesics,
so we identify the position and shape of the shocks using the same geodesic
equations described above. In particular, the shock front generators will jump
and deflect on collision at $u=v=0$ with the other shock and, for homogeneous
beams, will focus at $v_F = {1\over\b}$ and $u_F={1\over\a}$ for shocks 1 and 2 
respectively. At first sight, it may seem that our approximation of neglecting
the shock front distortion would be good provided the distortion is small, 
i.e.~the signal trajectory lies far from the focusing zone. This would 
require $u_Q \ll u_F$ and $v_S \ll v_F$. In terms of the parameters $\a,\b,a,b$
we find
\eqnn\sectdt
$$\eqalignno{
u_Q~~&=~~ u_0 ~~=~~{1\over\a}\Bigl(1-{b\over a}\Bigr) \cr
v_S~~&=~~ v_0 ~~=~~\a a b \cr
{}&{}& \sectdt \cr }
$$
The far-from-focus conditions are therefore:
\eqnn\sectdu
$$\eqalignno{
u_Q &\ll u_F ~~~~~~\leftrightarrow~~~~~~1-{b\over a} ~\ll~ 1 \cr
v_S &\ll v_F ~~~~~~\leftrightarrow~~~~~~\a \b a b ~~\ll~ 1 \cr
{}&{}& \sectdu \cr }
$$
the second being equivalent to $\tan{\varphi_a\over2} \tan{\varphi_b\over2} \ll 1$.

Now notice the curious fact that these conditions involve precisely the two
coefficients in the time machine criterion \sectdr. The magnitude of the
back-in-time motion S $\rightarrow$ R is therefore constrained to be small
by the far-from-focus condition. Moreover, the fact that the same parameters
are controlling both effects strongly suggests we need to look more closely
at the assumption that shock front distortion can be neglected.

In the non-interacting idealisation, the shock fronts propagate both before and after 
the collision as flat discs moving at the speed of light. The collision causes
the shock front generators to jump and deflect, so that for homogeneous beams
the wavefront shape of, for example, shock 2 after the collision is as shown in
Fig.~9.
 
\vskip0.2cm
{\epsfxsize=8cm\epsfbox{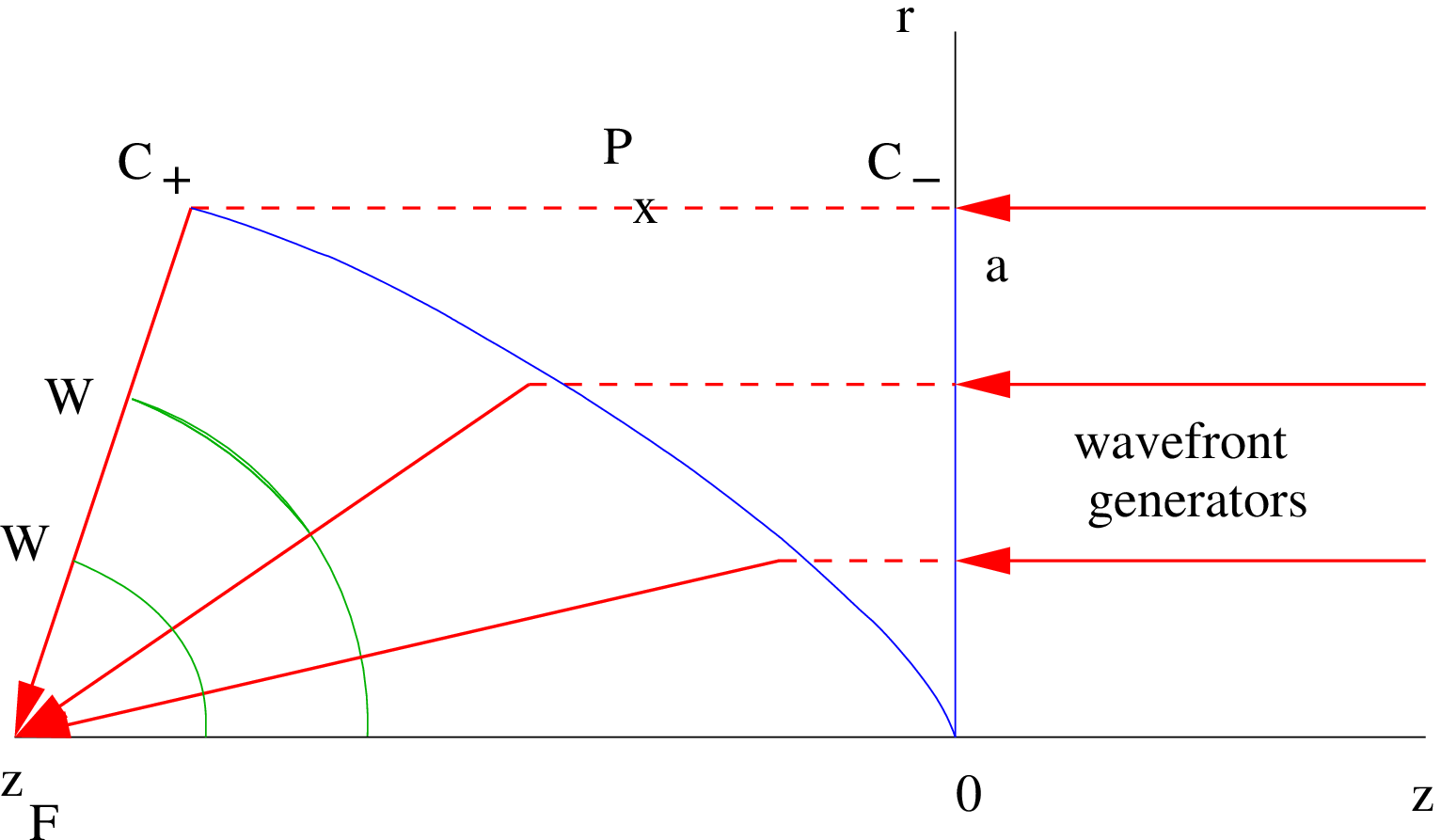}} 
\vskip0.01cm
\noindent{\eightpoint Fig.~9~~The shape of the gravitational wave shockfront for 
shock 2 after impact at the surface $C_-$ ($z=0$) with shock 1. The wavefront 
experiences a discontinuous jump in the negative $z$ direction (and backwards 
in $t$) to the surface $C_+$. It then focuses to the point $r=0$, $z_F = {1\over\a}$.
The equal-time wavefronts are the spherical surfaces $W$. See below for the
significance of point P.}
\vskip0.1cm

\vskip0.2cm
{\epsfxsize=5cm\epsfbox{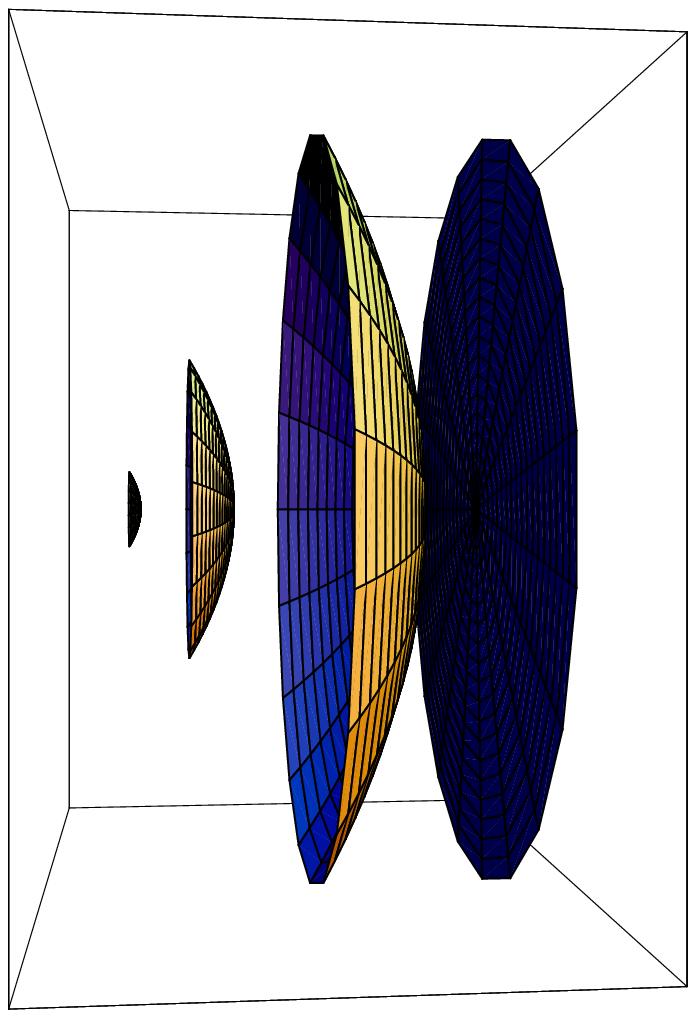}} 
\vskip0.01cm
\noindent{\eightpoint Fig.~10~~3D plot in $(z,x,y)$ coordinates of the equal-time 
shock wavefronts described in Fig.~8.}
\vskip0.1cm
The shock wavefronts, defined as equal $t$ surfaces, are spherical shells converging 
at the speed of light to the focal point $r=0,~ z=z_F={1\over\a}$. The relevant 
geodesic equations for their generators after the jump are
\eqnn\sectdv
$$\eqalignno{
v~~&=~~-\a a^2 (1-\a u) \cr
r ~~&=~~a (1-\a u) \cr
{}&{}& \sectdv \cr }
$$
and it is straightforward to show that the wavefronts are the surfaces
\eqn\sectdw{
r^2 ~~=~~-{1\over\a} v (1-\a u)~~~~~~~~~({\rm wavefronts})
}

The key observation, however, concerns the status of the after-jump surface
$C_+$. This is easily identified as the surface 
\eqn\sectdx{
z = -{1\over2} \a r^2 ~~~~~~~~~~({\rm collision~ surface}~ C_+)
}
This is the `collision surface' regarded as the limit $u \rightarrow 0_+$
of the shock fronts, which must be distinguished in these coordinates from the
`collision surface' $C_-$ identified as the limit $u\rightarrow 0_-$ of the
shock fronts from the pre-collision region. The important point is that the 
curved spacetime zone is in fact the 
future region of the collision surface $C_+$, {\it not} the `initial' $u=v=0$ 
collision disc $C_-$. 

In particular, this means that any point P behind the collision surface $C_{+}$
(see Fig.~9) at a later time is in fact in the curved 
zone\foot{This is especially evident, indeed trivial, if we use different coordinates. 
Defining the discontinuous change of coordinates to $\hat u$, $\hat v$, $\hat r$ with
$$\eqalignno{
u &= \hat u \cr
v &= \hat v + \theta (\hat u) \Bigl(f(\hat r) + {1\over4} f'(\hat r)^2 
\hat u \Bigr) \cr
r &= \hat r \Bigl(1 + \theta(\hat u) {1\over2}f'(\hat r) {1\over \hat r} \hat u
\Bigr) \cr}
$$
then by construction the geodesics expressed in hatted coordinates are
continuous across the shock. The shock fronts now remain
flat in these coordinates but the spacetime metric is non-trivial everywhere
and its Minkowski nature, which allows us some intuition as to
the meaning of the `time' coordinate, is hidden. (This is why we chose to analyse
the `time machine' trajectory in the original Aichelburg-Sexl coordinates.)
The spacetime now looks like the illustration in Fig.~8 in that the
shocks $\hat u =0$ and $\hat v = 0$ divide the spacetime into the four regions
I, $\ldots$ IV. Now, however, the incoming signal geodesic simply passes straight
through the $\hat u = 0$ shock wave and directly into region IV, which is
the curved zone. The collision surfaces $C_-$ and $C_+$, which must be carefully
distinguished in Aichelburg-Sexl coordinates, are clearly degenerate in the
hatted description.}. 
It is now reasonably clear to see that the point P on the time machine trajectory 
SPQR is in fact in this region. Comparing the point S with impact parameter $a$,
corresponding to the collision of the signal and shock 1, with the collision
at $u=v=0$ of the shock 2 generator with $r=a$ and shock 1, we see that both are
jumped back in time along the $u=0$ line by the {\it same} amount $f_1(a)$. The
point P is therefore necessarily to the future, with greater $z$, than the
shock 2 wavefront, as illustrated in Fig.~9. This is perhaps best seen using
the 3D plot in Fig.~7, where we can interpret the shaded surface as the
wavefront of shock 2; the signal at P would then lie on the flat sloping surface,
always {\it behind} (greater $t$, $z$ and $v$) the shock front. It is easy to
check that this also holds for any angle of incidence of the signal.

The resolution of the `time machine' scenario is therefore simply that the point Q
illustrated in Fig.~8 does {\it not} in fact represent a collision of the signal
with shock 2. Because the shock front has been distorted by the collision, the 
signal is now behind this curved shell and never catches up. Moreover, because it
is in the future of the `collison surface' $C_+$ (though not $C_-$) it is in
the curved zone where the full metric is unknown and we cannot describe its path
using the geodesic equations above.

Ultimately, therefore, this proposal for a time machine trajectory fails when
the interaction between the gravitational shock waves is taken correctly into
account. The shock fronts distort in the collision and tend to focus, with the
curvature of the shock 2 front sufficient by construction to keep the signal
trajectory behind it, in the future curved spacetime zone. 

Is there a way to evade this conclusion? We may try, for example, to let the
two finite-size shocks pass each other with a bigger impact parameter than
their sizes $R_1,R_2$ (regions with non-zero energy density $\r(r)$). But this 
does not help. The shock front is defined by the area of transverse space where
the profile function $f(r)$ is non-zero and this extends beyond the radii of
the shocks. Necessarily the initial signal collision with shock 1 takes
place at an impact parameter where the profile of shock 2 is non-zero and, assuming
the shock front is governed by the kinematics of the congruence of null geodesics
which generate it, necessarily they both experience a back-in-time jump by the 
same amount. The situation is therefore exactly as above. The only way to recover
the time-machine scenario of non-interacting shocks would be if the profile functions
vanished outside some radius. But since in particular $f'(r) = -8 E(r)/r$ where $E(r)$
is the total beam energy within the radius $r$, $f(r)$ can never vanish.
This would imply screening of the gravitational source. Such an `anti-gravity' 
theory would presumably have even more severe conceptual problems than those provided
by CTCs!

\vfill\eject

\newsec{A Time Machine from Superluminal Photons in Gravitational Fields?}

In this section, we discuss a proposal by Dolgov and Novikov \refs{\DN} to create
a time machine by exploiting the phenomenon of superluminal photon propagation
in gravitational fields originally discovered by Drummond and Hathrell 
\refs{\DH}. As in the last two sections, this time machine again
involves two gravitating objects in relative motion, but here the essential
feature giving rise to backwards-in-time motion is not the spacetime itself
but the assumed superluminal propagation of signals in the background 
gravitational field.

We begin with a very brief review of the Drummond-Hathrell mechanism \refs{\DH}
(see also refs.\refs{\Sthree,\Sfive}). This exploits the fact that if the  
strong equivalence principle (SEP) is violated by direct couplings of the
electromagnetic field to the curvature, then photons are no longer constrained
to follow null geodesics and may, depending on their polarisation and direction,
propagate along spacelike curves. The simplest form of such SEP-violating
interactions is given by the Drummond-Hathrell effective action
\eqn\sectea{
\C = \int dx \sqrt{-g}\biggl[
-{1\over4}F_{\m\n}F^{\m\n} + {1\over m^2}\biggl(
a R F_{\m\n}F^{\m\n} + b R_{\m\n} F^{\m\l} F^\n{}_\l
+ c R_{\m\n\l\r} F^{\m\n} F^{\l\r} \biggr)\biggr]
}
This action is induced by vacuum polarisation in QED in curved spacetime (in which 
case the couplings $a$, $b$ and $c$ are $O(\a)$ and $m$ is the electron mass), 
where it is valid in a low-frequency approximation where terms involving 
derivatives of the field strengths or curvature have been neglected. 
It remains controversial whether or not the characteristics of photon 
propagation in QED are changed by dispersion. An argument proposed by
Khriplovich \refs{\Khrip} suggests that there is no dispersion. However,
we have recently constructed an extended effective action \refs{\Ssix}
valid to all orders in the derivative expansion and have used this to deduce 
the dispersive properties of the Drummond-Hathrell effect \refs{\Sfive}. Apart 
from certain interesting exceptional cases, the photon velocity was shown to 
be frequency-dependent, with the quantum vacuum in a gravitational field
acting as a dispersive medium. Unfortunately, this work was still not 
sufficiently general to give a definitive result for the high-frequency limit 
of the phase velocity, which determines the characteristics and thus the causal 
properties of photon propagation.
 
For our purposes here we shall set aside the issue of dispersion in QED,
and instead simply consider the Drummond-Hathrell action \sectea ~(with 
arbitrary, but small, couplings) as a phenomenological model of SEP violation,
which predicts non-dispersive, superluminal signal propagation. Within
this model, we want to address the question of whether superluminal
propagation necessarily leads to violations of causality and permits the 
construction of time machines. This is rather different from the previous
examples in that here we are concerned with whether there are physical
light paths that return to their temporal origin, even if the geometrical
structure of the background spacetime obeys standard causality properties.

The effective light cone derived from \sectea is \refs{\Sthree,\Sfour} 
\eqn\secteb{
k^2 ~+~{(2b+4c)\over m^2} R_{\m\l} k^\m k^\l ~-~{8c\over m^2} 
C_{\m\n\l\r} k^\m k^\l a^\n a^\r ~~\equiv~~ 
{\cal G}^{\m\n}(R,a) k_\m k_\n ~~=~~0
}
where $k_\m$ is the wave-vector and $a^\m$ the polarisation (normalised
so that $a^2=-1$). For vacuum spacetimes, only the term depending on the 
Weyl tensor is present and it is this that gives rise to the polarisation
dependence of the speed of light (gravitational birefringence). Since in this
model the modified light cone is still quadratic in $k_\m$, it can be
written as shown in terms of ${\cal G}^{\m\n}(R,a)$, which depends on the 
polarisation and local spacetime curvature.\foot{Notice that this is {\it not} 
the same relation as for conventional `tachyons', which have an inhomogeneous 
term in the mass-shell relation, i.e. $k^2 = -M^2$ in a metric with signature 
$(+---)$. The speed of such a tachyon is not fixed but depends on its energy.} 
The equivalent photon momentum 
$p^\m = {dx^\m\over ds}$, which is a tangent vector to the light rays 
$x^\m(s)$, is related non-trivially to the wave-vector by \refs{\LSV,\Sfive}
\eqn\sectebb{
p^\m = {\cal G}^{\m\n}k_\n    
}
With this definition, the photon momentum satisfies
\eqn\sectebbb{
G_{\m\n}p^\m p^\n ~~=~~ {\cal G}^{\m\n} k_\m k_\n~~=~~0
}
where $G\equiv {\cal G}^{-1}$ defines a new effective metric which determines 
the light cones mapped out by the light rays in the geometrical optics
construction. Notice that indices are always raised or lowered using the
original geometric metric $g_{\m\n}$, so that we need to distinguish
explicitly between the effective metric $G$ and its inverse ${\cal G}$.

It is elegant and instructive to re-express this condition using the Newman-Penrose
formalism (see, e.g.~ref.\refs{\Ch} for a review). 
We introduce a null tetrad in terms of the null vectors $\ell^\m$,
$n^\m$ and the complex null vectors $m^\m$ and $\bar m^\m$, where 
$m^\m = {1\over \sqrt2}(a^\m + i b^\m)$ with $a^\m$ and $b^\m$ spacelike.
These are chosen to satisfy the orthogonality and 
normalisation conditions
\eqn\sectec{
\ell.m ~=~ \ell.\bar m ~=~ n.m ~=~ n.\bar m  ~=~ 0~~~~~~
\ell.n ~=~1  ~~~~~~ m.\bar m ~=~ -1
}
Components of the Weyl tensor in the corresponding null tetrad basis are
denoted by the complex scalars $\Psi_0, \ldots, \Psi_5$, where
\eqnn\sected
$$\eqalignno{
\Psi_0 ~&=~ - C_{abcd}\ell^a m^b \ell^c m^d ~~~&=~~
-C_{1313} ~~~~~~~~~~~~~~~~~~~~~ \cr
\Psi_1 ~&=~ - C_{abcd}\ell^a n^b \ell^c m^d ~~~&=~~
-C_{1213} ~~~~~~~~~~~~~~~~~~~~~ \cr
\Psi_2 ~&=~ - C_{abcd}\ell^a m^b \bar m^c n^d ~~~&=~~
-C_{1342} ~~~~~~~~~~~~~~~~~~~~~ \cr
\Psi_3 ~&=~ - C_{abcd}\ell^a n^b \bar m^c n^d ~~~&=~~
-C_{1242} ~~~~~~~~~~~~~~~~~~~~~ \cr
\Psi_4 ~&=~ - C_{abcd}n^a \bar m^b n^c \bar m^d ~~~&=~~
-C_{2424} ~~~~~~~~~~~~~~~~~~~~~ \cr
{}&{}& \sected \cr }
$$
In this formalism, the modified light cone condition takes a simple form.
If we choose the unperturbed photon momentum to be in the direction of
the null vector $\ell^\m$, i.e.~$k^\m=\omega\ell^\m$, and the transverse 
polarisation vectors as $a^\m, b^\m$, then the light cone condition becomes 
(for Ricci-flat spacetimes)
\eqn\sectee{
k^2 ~~=~~\pm~{4c\omega^2\over m^2} (\Psi_0 + \Psi_0^*)
}
for the two polarisations respectively. It is clear from \sectee ~that
in Ricci flat spacetimes, if one photon polarisation travels subluminally,
the other necessarily has a superluminal velocity. Physical photons no
longer follow the geometrical light cones, but instead propagate on the
family of effective, polarisation-dependent, light cones defined by
eq.\sectee. 

Many examples of the Drummond-Hathrell effect in a variety of gravitational
wave, black hole and cosmological spacetimes have been 
studied \refs{\DH,\Sone,\Stwo,\Sfour} etc.
The black hole cases are particularly interesting, and it is
found that for photons propagating orbitally, the light cone is generically
modified and superluminal propagation occurs. For radial geodesics in
Schwarzschild spacetime, however, and the corresponding principal null geodesics
in Reissner-Nordstr\"om and Kerr, the light cone is unchanged. The reason is
simply that if we choose the standard Newman-Penrose tetrad in which
$\ell^\m$ is the principal null geodesic, the only non-vanishing component
of the Weyl tensor is $\Psi_2$ since these black hole spacetimes are all
Petrov type D, whereas the modification to the light cone condition
involves only $\Psi_0$.

Our principal interest here is in whether the existence of this type
of superluminal propagation in gravitational fields, which after all does 
imply backwards in time motion in a class of local inertial frames, is 
necessarily in contradiction with our standard ideas on causality. 
A precise formulation of the question is whether a spacetime which is 
{\it stably causal} with respect to the original metric remains stably causal 
with respect to the effective metric defined by the modified light cones 
\sectee.\foot{See also ref.\refs{\LSV} for a closely 
related discussion in the context of superluminal propagation in flat
spacetime between Casimir plates.}
A precise definition of stable causality can be found in 
ref.\refs{\Hawkone}, proposition 6.4.9. Essentially, this means that the 
spacetime should still admit a foliation into a set of hypersurfaces which
are spacelike according to the effective light cones.  

The Drummond-Hathrell effective light cones differ from the geometrical ones
by perturbatively small shifts dependent on the local curvature. While it is
certainly possible that these shifts are sufficient to destroy stable causality,
we have not identified any general argument that says this {\it must} be so.
Rather, it seems entirely possible that a curved spacetime could remain
stably causal with respect to the effective light cones. The issue fundamentally
comes down to global, topological properties of spacetime, such as are addressed
in ref.\refs{\Hawktwo}, but which lie rather beyond the scope of this paper.
Instead, we shall take a more pragmatic approach and simply try, following
ref.\refs{\DN}, to identify a time-machine scenario using superluminal
propagation.

The Dolgov-Novikov time machine is most simply (though not essentially)
described for a spacetime which admits superluminal radial trajectories.
Rather than use a black hole spacetime as suggested in ref.\refs{\DN}, we 
therefore describe the time machine in terms of the asymptotic spacetime
surrounding an isolated source of gravitational radiation,
which is described by the Bondi-Sachs metric. The Drummond-Hathrell effect in 
this spacetime has recently been studied in ref.\refs{\Sfour}.

The Bondi-Sachs metric is \refs{\Bondi,\Sachs}
\eqn\sectef{
ds^2 = -W du^2 - 2 e^{2\b} du dr + r^2 h_{ij}(dx^i - U^i du)
(dx^j - U^j du)
}
where
\eqn\secteg{
h_{ij}dx^i dx^j = {1\over2}(e^{2\c} + e^{2\d}) d\theta^2
+ 2 \sinh(\c - \d) \sin\theta d\theta d\phi
+ {1\over2}(e^{-2\c} + e^{-2\d}) \sin^2\theta d\phi^2
}
The metric is valid in the vicinity of future null infinity ${\cal I}^+$.
The family of hypersurfaces $u = const$ are null, i.e. $g^{\m\n}
\pl_\m u \pl_\n u = 0$ and their normal vector $\ell_\m$ satisfies
\eqn\secteh{
\ell_\m = \pl_\m u ~~~~~~~~~~~~~\Rightarrow~~~~~
\ell^2 = 0, ~~~~~~~~\ell^\m D_\mu \ell^\n = 0
}
The curves with tangent vector $\ell^\m$ are therefore
null geodesics; the coordinate $r$ is a radial parameter along these rays  
and is identified as the luminosity distance. Outgoing radial null geodesics
are associated with $\ell^\m$ while incoming null geodesics have tangent vector
$n^\m$ in the obvious Newman-Penrose basis \refs{\Sachs,\Sfour}. 
The metric functions have asymptotic expansions near ${\cal I}^+$ for 
large luminosity distance $r$. We just need those for the functions
$\c\pm\d$, which correspond to the two independent gravitational wave
polarisations:
\eqnn\sectei
$$\eqalignno{
{1\over2} (\c + \d) &= {c_+ \over r}
+ {q_+ \over r^3}+ O({1\over r^4})\cr
{1\over2} (\c - \d) &= {c_\times \over r}
+ {q_\times \over r^3}+ O({1\over r^4})\cr
{}&{}& \sectei\cr
}
$$
where $c_{+(\times)}$  and $q_{+(\times)}$ are functions of $(u,\theta,\phi)$.
$\pl_u c_{+(\times)}$ are the Bondi `news functions', while 
$\pl_u^2 c_{+(\times)}$
can be identifed as the gravitational wave amplitude in a weak-field limit.
$q_{+(\times)}$ denote the `quadrupole aspect' of the gravitational radiation.

As shown in ref.\refs{\Sfour}, the modifications to the light cone following
from eq.\secteb ~are as follows. For outgoing photons, with $k^\m = \omega\ell^\m$,
\eqn\sectej{
k^2 ~~=~~\pm{4c\omega^2\over m^2}~ \bigl(\Psi_0 + \Psi_0^*\bigr)~~
=~~\pm {48c\omega^2\over m^2} {q_+\over r^5}
}
for the two photon polarisations aligned with the $+$ gravitational wave 
polarisation. A similar result holds for alignment with the $\times$ 
polarisation. For incoming photons,
\eqn\sectek{
k^2 ~~=~~\pm{4c\omega^2\over m^2}~ \bigl(\Psi_4+ \Psi_4^*\bigr)~~
=~~\pm {8c\omega^2\over m^2}~{1\over r}\pl_u^2 c_+ 
}
The different dependence on $1/r$ is governed by the Peeling Theorem in
Bondi-Sachs spacetime \refs{\Sachs}, according to which the asymptotic form 
of the Weyl tensor components is $\Psi_4 \sim {1\over r}$, $\Psi_3 \sim 
{1\over r^2}$, $\ldots$, $\Psi_0 \sim {1\over r^5}$.

Depending on the photon (and gravitational wave) polarisations, we therefore
have superluminal propagation for both outgoing and incoming radial photons,
albeit only of $O({1\over r^5})$ for the photons in the same direction as
the gravitational waves. The Bondi-Sachs spacetime therefore provides
a natural realisation of the Dolgov-Novikov (DN) scenario.\foot{Another simple
possibility with superluminal radial trajectories would be the dilaton
gravity model considered in ref.\refs{\Cho}.}

The DN proposal is to consider first a gravitating source with a superluminal
photon following a trajectory which we may take to be radial. Along this 
trajectory, the metric interval is effectively two-dimensional and DN
consider the form
\eqn\sectel{
ds^2 =  A^2(r) dt^2 - B^2(r) dr^2
}
(Actually, in order to realise the Bondi-Sachs case, we would need to let the
coefficients $A$ and $B$ depend on both $r$ and $t$ because of the
time-dependence of the gravitational wave metric. This complication is 
inessential for the main ideas, so we neglect it and simply follow the DN 
calculation based on eq.\sectel.) The photon velocity in the $(t,r)$ coordinates
is taken to be $v = 1 + \d v$, so the effective light cones lie perturbatively
close to the geometric ones. The trajectory is forward in time with respect 
to $t$.

DN now make a coordinate transformation corresponding to a frame in relative
motion to the gravitating source, rewriting the metric interval along the
trajectory as
\eqn\sectem{
ds^2 = A^2(t',r') \bigl(dt'^2 - dr'^2\bigr)
}
The transformation is
\eqnn\secten
$$\eqalignno{
t' &= \c(u)\bigl(t - ur -u f(r) \bigr) \cr
r' &= \c(u)\bigl(r - ut + f(r) \bigr) \cr
{}&{}& \secten\cr
}
$$
with 
\eqn\secteo{
f(r) = \int dr~\Bigl({B\over A} - 1\Bigr)
}
This is best understood in two steps. First, since any two-dimensional metric
is conformally flat, we can always (even if $A$ and $B$ were functions of both
$r$ and $t$) find a transformation $(t,r) \rta (\tilde t, \tilde r)$ which
takes the metric to the standard form $ds^2 = \Omega^2
\bigl(d\tilde t^2 - d\tilde r^2\bigr)$. For \sectel, the transformation to 
standard conformally flat form is achieved by
\eqnn\sectep
$$\eqalignno{
\tilde t &= t \cr
\tilde r &= r + f(r) \cr
{}&{}& \sectep\cr
}
$$
with $f(r)$ as in \secteo. The second step is just a boost on the flat
coordinates $(\tilde t, \tilde r)$, viz.
\eqnn\secteq
$$\eqalignno{
t' &= \c(u)\bigl(\tilde t - u \tilde r \bigr) \cr
r' &= \c(u)\bigl(\tilde r - u \tilde t \bigr) \cr
{}&{}& \secteq\cr
}
$$

Now, a superluminal signal with velocity 
\eqn\secter{
v = 1 + \d v = {B\over A} {dr\over dt}
}
emitted at $(t_1,r_1)$ and received at $(t_2,r_2)$ travels forward in $t$
time (for small, positive $\d v$) with interval
\eqn\sectes{
t_2 - t_1 = \int_{r_1}^{r_2} dr ~\bigl(1- \d v\bigr) {B\over A}
}
As DN show, however, this motion is backwards in $t'$ time for sufficiently
large $u$, since the equivalent interval is
\eqn\sectet{
t'_2 - t'_1 = \c(u)~\int_{r_1}^{r_2} dr ~\bigl(1- u -\d v\bigr) {B\over A}
}
The required frame velocity is $u > 1 - \d v$, i.e.~since $\d v$ is small,
$u > {1\over v}$.

The situation so far is therefore identical in principle to the discussion of 
superluminal propagation illustrated in section 2, Fig.~1. In DN coordinates 
the outward superluminal signal is certainly propagating backwards in time, but a 
reverse path with the same perturbatively superluminal velocity would necessarily
go sufficiently forwards in time to arrive back within the future light cone
of the emitter.\foot{
Notice that in the Bondi-Sachs case, the return signal could also be arranged
to have a superluminal velocity (of $O({1\over r})$ rather than the outward
$O({1\over r^5})$~) assuming the polarisation can be reversed without altering
any other features of the kinematics.}

As yet, therefore, we have no time machine but have simply recovered standard 
features of superluminal motion described in different frames. At this point,
DN propose to introduce a second gravitating source moving relative to the
first, as illustrated below.
\vskip0.2cm
{\epsfxsize=9cm\epsfbox{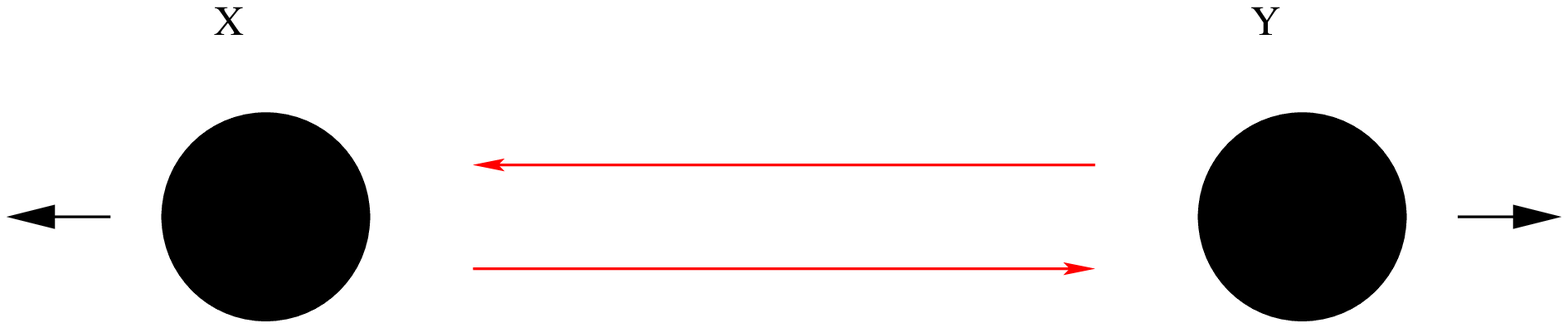}} 
\vskip0.1cm
\noindent{\eightpoint Fig.~11~~The Dolgov-Novikov time machine proposal. A
superluminal signal from $X$, described as backwards-in-time in a relevant 
frame, is sent towards a second gravitating source $Y$ moving
relative to $X$ and returned symmetrically.}
\vskip0.1cm
\noindent This is reminiscent of the cosmic string and shock wave proposals
in that a single gravitational source provides the essential physical
effect but two sources in relative motion are required to construct a genuine
time machine scenario.

DN now claim that a superluminal photon emitted with velocity $v(r)$ in the
region of $X$ will travel backwards in time (according to the physically
relevant coordinate $t'$) to a receiver in the region of $Y$. A signal
is then returned symmetrically
to be received at its original position in the vicinity of $X$, arriving,
according to DN, in its past. This would then be analogous to the 
situation illustrated in Fig.~2.  

However, as we emphasised in section 2, we are {\it not} free to realise the
scenario of Fig.~2 in the gravitational case, because the SEP-violating 
superluminal propagation proposed by Drummond and Hathrell is pre-determined, 
fixed by the local curvature. The $t'$ frame may describe back-in-time 
motion for the outward leg, but it does not follow that the return path
is similarly back-in-time {\it in the same frame.} The appropriate special 
relativistic analogue is the scenario of Fig.~1, not Fig.~2. This critique of 
the DN time machine proposal has already been made by Konstantinov \refs{\Konst} 
and further discussion of the related effect in flat spacetime with Casimir 
plates is given in \refs{\LSV}. The relative motion of the two sources, which 
at first sight seems to allow the backwards-in-time coordinate $t'$ to be 
relevant and to be used symmetrically, does not in fact alleviate the problem.
 
The true situation seems rather more to resemble the following figure.
\vskip0.2cm
{\epsfxsize=9cm\epsfbox{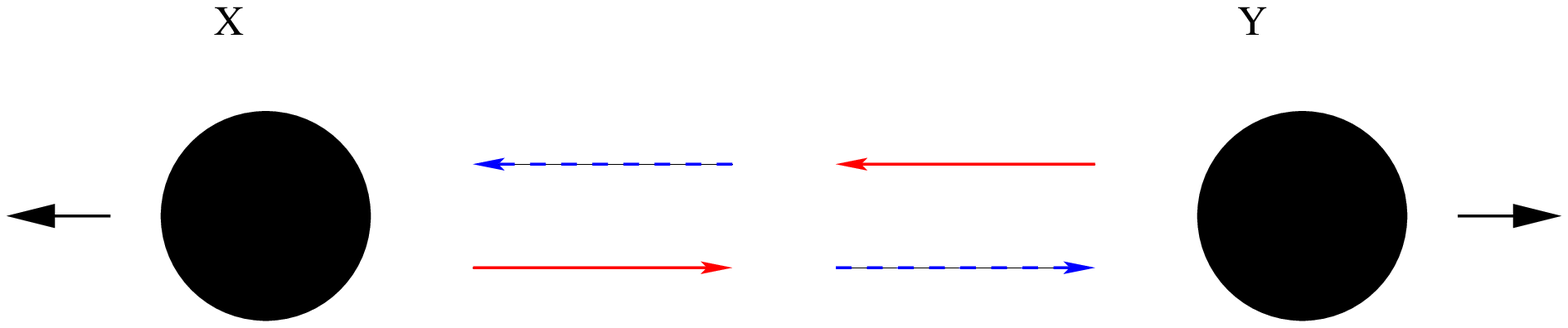}} 
\vskip0.01cm
\noindent{\eightpoint Fig.~12~~A decomposition of the paths in Fig.~11 for
well-separated sources.}
\vskip0.1cm
With the gravitating sources $X$ and $Y$ sufficiently distant that spacetime
is separated into regions where it is permissible to neglect one or the other,
a signal sent from the vicinity of $X$ towards $Y$ and back would follow
the paths shown. But it is clear that this is no more than stitching together
an outward plus inward leg near source $X$ with an inward plus outward leg
near $Y$. Since both of these are future-directed motions, in the sense of
Fig.~1, their combination cannot produce a causality-violating trajectory.
If, on the other hand, we consider $X$ and $Y$ to be sufficiently close that
this picture breaks down, we lose our ability to analyse the Drummond-Hathrell
effect since the necessary exact solution of the Einstein equations describing 
the future zone of the collision of two Bondi-Sachs sources is unknown. 

We therefore conclude that the Dolgov-Novikov time machine, like the others
considered in this paper, ultimately does not work. The essential idea
of trying to realise the causality-violating special relativistic scenario 
of Fig.~2 by using two gravitational sources in relative motion does not 
in the end succeed, precisely because the physical Drummond-Hathrell light cones
\secteb ~are fixed by the local curvature. It appears, surprising though
it may appear at first sight, that in general relativity with SEP-violating
electromagnetic interactions, superluminal photon propagation and causality
may well be compatible.

\vskip1cm

\newsec{Conclusions}

In this paper, we have analysed three proposals for constructing time 
machines which, although quite different, share a common basic structure, 
and have shown how each of them is in fact ultimately
consistent with our fundamental idea of causality.

The first, the Gott cosmic string time machine, exploits the conical
nature of the cosmic string spacetime to find effectively superluminal
paths. Introducing a second cosmic string then leads to the existence
of CTCs encircling the two strings, provided the string velocities satisfy 
the condition $u > \cos\a$, where $2\a$ is the deficit angle related to the 
string mass. However, provided appropriate boundary conditions are imposed, 
global consistency of the two-string solution requires exactly the opposite 
condition on the velocities, $u < \cos\a$. This condition was formulated in the
language of holonomy for loops in the two-string spacetime. The intricate
question of whether these boundary conditions should be considered
physical and the relation of Lorentz boost holonomies to tachyonic motion
was also discussed.

The second example mimics the Gott proposal but this time exploits the
discontinuous geodesics which arise in gravitational shock wave spacetimes
as the mechanism for effectively superluminal paths. In the unrealistic
approximation of neglecting the interaction between the shocks, CTCs were
also found in which a test particle is struck successively by the two
shocks, the first collision reversing its direction of motion as well
as inducing a time shift. However, a self-consistent analysis shows that
because the shock fronts are themselves generated by a congruence of
null geodesics, they are also distorted by the same mechanism responsible
for the discontinuities in the test particle geodesics. The shock front
distortion is exactly sufficient to invalidate the CTC construction.
The motivation for this construction, namely that the reason for the
absence of CTCs when the shock wave interactions are included would shed
new light on the kinematics of event horizon formation in the collision,
was not realised in this case.

The final example exploited the genuinely superluminal propagation arising
from a SEP violating effective action to look for causality violations.
General arguments why this specific type of superluminal propagation
need not be in conflict with stable causality were discussed. A critical
analysis of the proposal by Dolgov and Novikov to use this mechanism
to build a time machine again involving two gravitational sources
in relative motion was then made, with the by now familiar conclusion
that while causal violations appear to arise when the two sources are
treated independently, a consistent analysis of the combined spacetime
shows that closed light trajectories do not occur.

We close with some general reflections on these 
diverse examples, and their relevance for future theories and theorems.
All of the scenarios discussed here rely fundamentally on finding effectively
superluminal paths, whether by constructing spacetimes with missing wedges
or non-trivially patched regions of Minkowski spacetime as in the cosmic string
and shock wave models, or by assuming a modification of electrodynamics in which
light itself no longer follows null paths. Both of the first two models fall
within the orbit of the chronology protection theorems, even though the Gott
spacetime is extremely subtle. The final example is different in that 
it breaks the assumption made in all conventional work on causality in general 
relativity that physical signals follow timelike or null geodesics. Now it may well
be true (see ref.\refs{\Sfive} for a careful recent discussion) that despite the
Drummond-Hathrell effect in QED, the `speed of light' relevant for causality,
i.e.~the high-frequency limit of the phase velocity, is indeed $c$. However,
this effect highlights the fact that theoretically consistent generalisations
of (quantum) electrodynamics can be constructed in which light signals no longer 
follow null geodesics and the physical light cones define a new metric which 
does not coincide with the geometric one. An elegant example of such a bi-metric 
theory was recently formulated in ref.\refs{\Dmetric}. It may well, therefore, be 
interesting to revisit the chronology protection theorems in these more 
general contexts. 

Finally, although the models described in this paper have been fundamentally
classical, the most interesting questions on the nature of causality arise
in quantum theory, where in particular the weak energy condition may be
circumvented. Although the indications from studies of quantum field theory
in curved spacetime suggest that the occurrence of CTCs leads to pathologies
such as divergent expectation values or violations of unitarity, perhaps
these are issues that also deserve renewed attention from the perspective of
modern ideas in quantum gravity. There are surely further scenarios to devise
which may yet successfully challenge Hawking's chronology protection 
conjecture \refs{\Hawktwo}: `{\it
the laws of physics do not allow the appearance of closed timelike curves}'.

\vskip1.5cm

\noindent{\bf Acknowledgments}

I am especially grateful to I.~Drummond, K.~Kunze, M.~Maggiore,
W.~Perkins and G.~Veneziano for helpful comments on these ideas,
and many others for interesting discussions on time machines.
Parts of this work were carried out while I held visiting positions
at CERN and the University of Geneva and I am grateful for their
hospitality. This research is also supported in part by PPARC grant 
PPA/G/O/2000/00448.

\vfill\eject

\listrefs

\bye